\documentclass[journal]{IEEEtran}

\usepackage{algorithmicx}
\usepackage[ruled]{algorithm}
\usepackage{algpseudocode}
\usepackage{cite}

\usepackage[english]{babel}
\usepackage[latin1]{inputenc}

\usepackage{epsfig}
\usepackage{color}

\usepackage{times}
\usepackage{subfigure}
\usepackage{url}
\usepackage{amsthm}
\usepackage{amsmath}
\usepackage{amssymb}
\usepackage{dsfont}

\theoremstyle{definition} 
\newtheorem{lemma}{Lemma} 
\newtheorem{theorem}{Theorem}
\newtheorem{remark}{Remark} 

\newcommand{\Ex}{\mathbb{E}}

\newcommand{\R}{\mathbb{R}}

\newcommand{\Hip}{\mathcal{H}}
\newcommand{\N}{\mathcal{N}}

\newcommand{\diag}{\text{diag}}
\newcommand{\tr}{\text{tr}}
\newcommand{\ve}[1]{\boldsymbol{#1}}

\newif\ifshowfig 
\showfigtrue     

\begin{document}

\title{Design and performance analysis of a fully distributed source detection algorithm for WSNs}

\author{{Juan Augusto Maya and Leonardo Rey Vega,~\IEEEmembership{Member,~IEEE,}}
\thanks{The two authors are with Universidad de Buenos Aires, Buenos Aires, Argentina. L. Rey Vega is also with CSC-CONICET, Buenos Aires, Argentina. 
Emails:  \{jmaya, lrey\}@fi.uba.ar. This work was partially supported by the projects UBACyT 20020170200283BA and PICT-2017-4533. This is a pre-print version of an article submitted to IEEE Transactions on Signal Processing and it is under review.} %
} 
\maketitle

\begin{abstract}
In this article, we consider the detection of a localized source emitting a signal using a wireless sensor network (WSN). We consider that geographically distributed sensor nodes obtain energy measurements and compute cooperatively and in a distributed manner a statistic to decide if the source is present or absent without the need of a central node or fusion center (FC). We first start from the continuous-time signal sensed by the nodes and obtain an equivalent discrete-time hypothesis testing problem. Secondly, we propose a fully distributed scheme, based on the well-known generalized likelihood ratio (GLR) test, which is suitable for a WSN, where resources such as energy and communication bandwidth are typically scarce. In third place, we consider the asymptotic performance of the proposed GLR test. The derived results provide an excellent matching with the scenario in which only a finite amount of measurements are available at each sensor node. We finally show that the proposed distributed algorithm performs as well as the global GLR test in the considered scenarios, requiring only a small number of communication exchanges between nodes and a limited knowledge about the network structure and its connectivity. 
\end{abstract}





\section{Introduction}
In the near past, Wireless Sensor Networks (WSN) have received considerable attention from the research and industrial community because of their remote monitoring and control capabilities \cite{shaikh2016energy, rashid2016applications, gungor2009industrial}. More recently, they have become an essential part of the emerging technology of Internet of Things (IoT) \cite{gubbi2013internet, al2015internet}. Among the different tasks to be done by WSNs, distributed detection is an actively researched topic \cite{chepuri2016sparse,ciuonzo2017distributed, al2018node}.

Distributed detection architectures can be broadly classified in two classes. In the first class all sensors transmit their local measurements to a fusion center (FC), where some processing tasks are done and the final decision about the underlying phenomenon is made \cite{tsitsiklis1989decentralized, blum1997distributed, viswanathan1997distributed, ChamberlandVeeravalli2003}. In many applications it is unfeasible or expensive to develop an infrastructure with a FC. This centralized architecture also presents some weaknesses as, for example, its lack of robustness against the malfunctioning of a single device, given that a failure in the FC may severely degrade the performance of the system. Additionally, it requires that sensor nodes, typically battery-powered devices, communicate through orthogonal channels with the FC, consuming excessive energy and bandwidth. A way to circumvent this issue is to quantize the measurements to few bits (binary quantization is a popular choice) to save bandwidth. However, this strategy involves the design of the quantizers, which can be a hard task when the observations are correlated typically resulting in complicated decision rules \cite{blum1996necessary,tenney1981detection} even under the assumption of Gaussian data and networks with only a few nodes \cite{willett2000good}.

An alternative to the above described architecture is to consider distributed strategies for which there is not a central processing unit or FC. In this type of detection architectures, sensors distributed geographically, collect measurements from the phenomenon of interest, make some processing, exchange information with their neighbors and, finally, execute some consensus or diffusion algorithm to achieve their respective decisions. This option is robust against node failures, and the communications between nodes are done locally, over typically short distances, saving energy and also bandwidth, by employing spatial reuse of the frequency bands. Thus, the quantization of the measurements can be done with more levels and it becomes a less relevant problem.  

Many works have considered the second option also known as a fully distributed detection architecture \cite{sayed2014adaptation,al2018node,sayed2013diffusion,kar2008topology}. Nevertheless, most of the work found in the literature assumes that the spatial measurements are independent or, they ignore the statistical dependence of the data when designing the distributed detection algorithms \cite{ciuonzo2014decision, hamed2012reliable, varshney1986optimal,kar2011distributed,braca2010asymptotic,li2018fully}. For example, Cattivelli et al. proposed in \cite{cattivelli2011distributed} a distributed detection algorithm to detect a known deterministic signal under Gaussian noise, where the noise is assumed to be independent across the sensors, and thus the observations at each node are independent under each hypothesis. However, in many applications of interest, the measurements taken by spatially distributed nodes are statistically dependent, and disregarding this effect markedly degrades the detection performance of the network \cite{drakopoulos1991optimum}. 


Other works have considered dependent observations using Gaussian Markov Random Fields \cite{Tong_2007, anandkumar2009scalable} to design a Neyman-Pearson detector in a centralized scenario. However, the design of distributed detection algorithms with dependent measurements in a decentralized scenario deserves more investigation \cite{javadi2016detection}. 

In this work we deal with spatially correlated observations and propose a fully distributed algorithm to detect the presence or absence of a localized source emitting a stochastic signal. This problem is important in its own with multiple applications in fields as cognitive radio \cite{Lunden_Koivunen_Poor_2015}, massive MIMO wireless networks \cite{Wang_Qiu_Zhao_2017} and acoustic source detection, separation and localization \cite{Wang_Reiss_Cavallaro_2016}, among others.

\subsection{Main contributions}
The main contributions of the work can be summarized as follows. First, we develop a model
for the problem of source detection. We assume that under the null hypothesis ($\Hip_0$) the signal is absent and under the alternative hypothesis ($\Hip_1$) it is present. The location of the source is unknown along with other parameters of the stochastic signal $s(t)$ that models the signal emitted by the source. Also, our network model does not assume the presence of a FC. The desired goal is the distributed detection of the source signal if present (that is, all the sensor nodes have to reach the same decision). Assuming that the nodes sense the energy of a signal, we are able to model the statistical dependence between samples in different nodes under both hypotheses. To make the problem tractable, we use the Central Limit Theorem to approximate the statistics of the observations by a multivariate Gaussian distribution under each hypothesis, where the covariance matrix under $\Hip_1$ has a particular structure that can be exploited to simplify the detection algorithm. 

Secondly, we compute a modified version of the generalized likelihood ratio (GLR) detector, which estimates the unknown parameters \emph{locally} at each node, instead of doing that \emph{globally}, which would consume more network resources and would require a distributed solution of a complex optimization problem. We also derive its asymptotic performance and prove that under mild conditions, it coincides with the asymptotic performance of the statistic that uses the global estimation.

In third place, we provide a fully distributed detector that can be efficiently computed using a spatial averaging algorithm where the communication between sensors is done locally and where the required prior knowledge at each sensor about the network connectivity is minimal. Its performance is evaluated using numerical simulations showing excellent results for a wide range of signal-to-noise ratios values.

\subsection{Organization}
The paper is organized as follows.  We present the detection problem and compute the statistics of the  measurements taken by the nodes in Section II. In Section III, we 
first propose to estimate the unknown network parameters at each node locally, and then, we compute the asymptotic distribution of this statistic under each hypothesis, which allows to characterize its asymptotic performance. In Section IV, we simplify this detector to one that can be efficiently implemented in WSNs via a consensus algorithm. In Section V, we evaluate the performance of the algorithm numerically and finally, in Section VI, we draw the main conclusions of this work. The proofs of some of the presented mathematical results are relegated to the appendices.

\subsection{Notation}

We will denote with $\ve 1_N$ the $N$-dimensional vector with all its entries equal to one, with $\mathbf{0}_N$ the $N$-dimensional null vector and with $\mathbf{I}_N$ the $N$-dimensional identity matrix. 
Given a vector $\mathbf{a}$ we denote with $ \text{diag}(\mathbf{a})$ a diagonal matrix with diagonal entries given by the components of $\mathbf{a}$. Similarly, given a square matrix $\mathbf{A}$, we denote with $\text{diag}(\mathbf{A})$ a diagonal matrix which preserves the diagonal of $\mathbf{A}$. With $ \N(\ve \mu,\ve \Sigma)$ we denote a multivariate normal distribution with mean vector $\ve\mu$ and covariance matrix $\ve\Sigma$. Given two $N$-dimensional vectors $\mathbf{a}$ and $\mathbf{b}$ we write $\mathbf{a}\succeq\mathbf{b}$ ($\mathbf{a}\succ\mathbf{b}$) if $a_i\geq b_i$ ($a_i>b_i$) for all $i\in[1:N]$.

\section{Model}
\label{sec:model}
\begin{figure}
\centering
\subfigure[][Netwotk topology.]{\includegraphics[width=.7\linewidth]{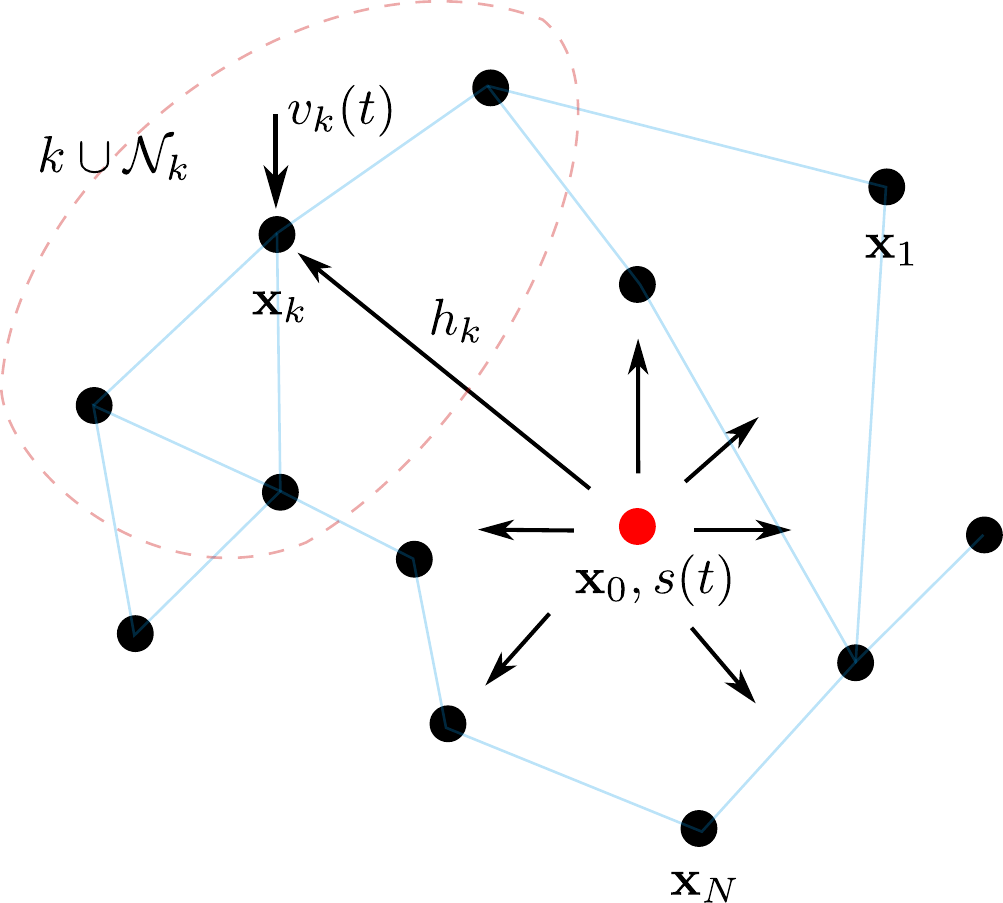}}
\subfigure[][Sensing node scheme.]{\includegraphics[width=\linewidth]{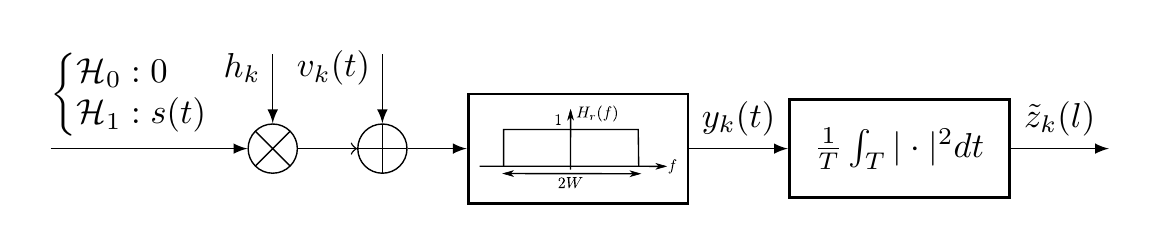}}
\caption{A scheme of the network is shown in (a) with nodes (black dots) sensing a source emitting a signal $s(t)$, which  propagates through channels with gain $h_k$. It is also shown the neighborhood of sensor $k$, noted $\N_k$ (see Section IV) and the sensing noise $v_k(t)$. A scheme of how each sensor processes the received signal is shown in (b).}
\label{fig:sensingnode}
\end{figure}

We consider a WSN with $N$ nodes with sensing capabilities distributed in a bounded geographical area. Each sensor position is denoted by $\mathbf{x}_k\in\mathbb{R}^2$ with $k\in[1:N]$. We will assume that at an unknown position $\mathbf{x}_0$ there is a possibility of having a source emitting a signal $s(t)$. Each sensor has a observation window of duration $\tau$ in which observes a signal $y_k(t)$, $t\in[0,\tau]$ $k\in[1:N]$. Through the processing of their observations the network looks for the correct decision regarding the presence or absence of the source in a fully distributed manner. This means that each sensor node has to reach the same decision about the presence or absence of the source without the help of a FC. This leads us to the following binary hypothesis testing problem
\begin{equation}
\left\{
\begin{array}{ll}
\!\!\!\Hip_0: y_k(t)=v_k(t), & \!\!\!t\in[0,\tau], k\in[1:N]\\
\!\!\!\Hip_1: y_k(t)=h_k s(t) 
+ v_k(t),&\!\!\! t\in[0,\tau], k\in[1:N]
\end{array}\right.
\label{eq:htp_cont}
\end{equation}
where $v_k(t)$ is a zero-mean base-band Gaussian complex circular\footnote{We will work with low-pass complex equivalent signals.} sensing noise with flat spectra with value $N_0$ and independent through the sensors. The source signal $s(t)$ is also assumed to be a zero-mean base-band Gaussian complex circular stationary stochastic process, independent from the sensing noise signals $v_k(t),\ k\in[1:N]$. It also assumed that the spectrum of $s(t)$ is again flat with value $N_s$ (this model can be generalized in several ways, see Remark \ref{remark:1} at the end of the section). We consider that the sensing system at each node has a limited two-sided bandwidth of $2W$, which leads us to stochastic signals with limited bandwidth under both hypothesis. The value of $h_k$ is assumed to be constant during the whole observation window and takes into account the characteristic of the wireless path between the source position $\mathbf{x}_0$ and the $k$-th sensor one, $\mathbf{x}_k$. It is in general a complex value which models attenuation and delay between the source position and node node $k$. For example, if we assume that a power-law attenuation is valid, then $|h_k|=\frac{1}{\epsilon+\|\mathbf{x}_k-\mathbf{x}_0\|^{\frac{\alpha}{2}}}$ where $\alpha$ is the path-loss exponent and $\epsilon$ is small constant. See in Fig. \ref{fig:sensingnode} a sketch of the network model and the sensing node scheme.

We will consider that each sensor has an energy detector being able to compute the energy of the received signals over some time window. The use of the energy detector is justified by practical considerations in a distributed setting, where a coherent detector at each sensor will require a very precise network-wide synchronization in order to take advantage of the phase or delay information of the measurements. Although some basic synchronization is always needed, energy detectors do not need such a fine synchronization as in the coherent case which could be expensive in terms of resources and difficult to achieve, specially in settings where the network have a large number of nodes. In more precise terms,  we assume that the full observation window of length $\tau$ can be divided in $L=\lfloor\frac{\tau}{T}\rfloor$ time slots with duration $T$. In each slot, each sensor uses its energy detector for computing:
\begin{equation*}
\tilde{z}_{k}(l)=\frac{1}{T}\int_{(l-1)T}^{lT}|y_k(t)|^2 dt,\ \ l\in[1:L],\ k\in[1:N].
\end{equation*}
As the signals under both hypotheses are bandwidth-limited and with flat spectrum we can use the Karhunen-Loève (KL) expansion for band-limited processes \cite{Papoulis_Pillai_2002}. More precisely, $y_k(t)=\sum_{p=1}^\infty a_{l,p,k} \phi_p(t-(l-1)T)$ under $\mathcal{H}_0$, and $y_k(t)=\sum_{p=1}^\infty (a_{l,p,k}+s_{l,p}) \phi_p(t-(l-1)T)$ under $\mathcal{H}_1$, for every $k\in[1:N]$, and for $t\in[(l-1)T,lT]$, where the \emph{eigenfunctions} $\left\{\phi_p(t)\right\}_{p=1}^\infty$ are orthonormal and complete
in $L_2\left([0,T]\right)$, the space of square-integrable functions, and satisfy the integral equation:
\[\int_{0}^T \frac{\sin{(2\pi W (s-T/2))}}{\pi (s-T/2)}\phi_p(s)ds=\lambda_p\phi_p(s),\]
where for each $p\in\mathbb{N}$, the \emph{eigenvalues} satisfies $\lambda_p> 0$.
The quantities $a_{l,p,k}$ and $s_{l,p}$ are complex and circular Gaussian independent random variables with zero mean and second order moment equal to $2\lambda_p N_0 W$ and $2\lambda_p N_s W$ respectively. If the time-bandwidth product verifies $2TW\equiv M\gg 1$ we can mimic the ideas in \cite{Urkowitz_1967}, and use the orthonormality of $\left\{\phi_p(t)\right\}_{p=1}^\infty$ to show that, for $k\in[1:N]$ and $\ l\in[1:L]$:
\begin{equation}
\left\{
\begin{array}{llll}
\Hip_0:& \tilde{z}_{k}(l) \approx\frac{1}{M}\sum^{M}_{p=1}|a_{l,p,k}|^2 \\ 
\Hip_1:& \tilde{z}_{k}(l) \approx\frac{1}{M}\sum^{M}_{p=1}|h_k s_{l,p}+a_{l,p,k}|^2.
\end{array}\right.
\label{eq:htp0}
\end{equation}
It is well-known \cite{Landau_Pollak_1962,Slepian_1965} that when  $M\gg 1$, $\lambda_p\approx 1$ when $p\in[1:M]$ and $\lambda_p\approx 0$ when $p>M$ which implies that $\mathbb{E}\left[|a_{l,p,k}|^2\right]\approx 2N_0W\equiv\sigma_v^2$ and $\mathbb{E}\left[|s_{l,p}|^2\right]\approx 2N_sW\equiv\sigma_s^2$ for all $k\in[1:N],\ l\in[1:L],\  p\in[1:M]$.

Using the above detailed properties about the random coefficients, it is straightforward to show that under $\Hip_0$, $\mathbb{E}\left[\tilde{z}_k(l)\right]=\sigma_v^2$ and $\text{Cov}\left[\tilde{z}_k(l),\tilde{z}_n(l)\right]=\frac{\sigma_v^4}{M} \delta_{kn}$, for all $l\in[1:L]$, where $\delta_{kn}$ is the Kronecker delta.

Similarly, under $\Hip_1$, we have $\mathbb{E}\left[\tilde{z}_k(l)\right]=\sigma_v^2(1 +c_k)$, 
and $\text{Cov}\left[\tilde{z}_k(l),\tilde{z}_n(l)\right]=\frac{\sigma_v^4}{M}(c_k c_n + (2 c_k +1)\delta_{kn})$, for all $l\in[1:L]$ and $k,n\in[1:N]$, where 
\begin{equation}
c_k\equiv\sigma^2_s|h_k|^2/\sigma^2_v> 0.\label{eq:ck}
\end{equation} 
We can group the measurements of all sensors in the time window $l$ in the vector $\tilde{\ve z}_l\equiv[\tilde{z}_{1}(l),\dots,\tilde{z}_{N}(l)]^T\in\R^N$ and define $\ve c = [c_1,\dots,c_N]^T\succ \ve 0_N$. 
At this point, and assuming again that $M$ is sufficiently large\footnote{We also assume that the temporal correlation between $s_{l,p}$ and $a_{l,p,k}$ in different time slots decreases sufficiently fast (mixing property \cite{Billingsley_1995}).}, we use the multidimensional Central Limit Theorem (CLT) \cite{Billingsley_1995} to show that
\begin{equation}
\left\{
\begin{array}{llll}
\Hip_0:& \ve \{\tilde{\ve z}_l\}_{l=1}^{L} \overset{iid}{\sim} \N(\sigma_v^2 \ve 1_N,\frac{\sigma_v^4}{M} \mathbf{I}_N) \\
\Hip_1:&  \{\tilde{\ve z}_l\}_{l=1}^{L} \overset{iid}{\sim}  \N(\tilde{\ve \mu}(\mathbf{c}),\tilde{\mathbf\Sigma}(\mathbf{c})),
\end{array}\right.
\label{eq:htp}
\end{equation}
where for easy reference we defined $\tilde{\ve \mu}(\mathbf{c})\equiv \sigma_v^2(\ve c + \ve 1_N)$ and $\tilde{\mathbf\Sigma}(\ve c)\equiv \frac{\sigma^4_v}{M} \left(\ve c \ve c^T + 2 \text{diag}(\ve c)  + \mathbf{I}_N\right)$.
For further notational simplicity we will work on the following equivalent test obtained from (\ref{eq:htp}) using the change of variables: $\ve{z}_l = (\tilde{\ve z}_l- \sigma_v^2 \ve 1_N)/(\sigma^2_v/\sqrt{M})$.  
\begin{equation}
\left\{
\begin{array}{llll}
\Hip_0:& \ve \{\ve z_l\}_{l=1}^{L} \overset{iid}{\sim} \N(\ve{0},\mathbf{I}_N) \equiv p(\ve z_l; \ve\theta=\ve 0_N) \\
\Hip_1:&  \{\ve z_l\}_{l=1}^{L} \overset{iid}{\sim}  \N(\ve\mu(\ve c),\mathbf \Sigma(\ve c))\equiv p(\ve z_l; \ve\theta= \ve c),
\end{array}\right.
\label{eq:htpeq}
\end{equation}
where 
\begin{equation}
\begin{array}{ll}
\ve \mu(\ve c) &= \sqrt{M}\ve c\\
\mathbf \Sigma(\ve c) & = \ve c \ve c^T + 2 \text{diag}(\ve c)  + \mathbf{I}_N.
\end{array}
\label{eq:meanCoveq}
\end{equation}
We will consider that the noise variance $\sigma_v^2$ is known (or estimated) at each node, 
and that $\ve c$ is unknown. This is because of the lack of knowledge of the true position of the source, but also due to the fact that the exact nature of the wireless path between each sensor is not known exactly (e.g. the value of the path-loss parameter $\alpha$) and it is also influenced by several complex phenomena (e.g. shadowing, fading, etc) which are difficult to know and model in advance. Therefore, we need a statistic that avoids the use of the unknown parameters or estimates them in some way. We will attack this issue in the next section.

It is important to observe that the vectors $\ve z_l$ contains the measurements taken in each sensor at the time slot $l\in[1:L]$. As we do not assume the presence of a FC and the sensors are geographically separated, we need to allow cooperation and communication between them in order to obtain a common and distributed decision about the presence or absence of the source. More precisely, we assume that sensors can communicate through error-free channels with other sensors in their neighborhood. The concept of neighborhood is naturally introduced modeling the WSN as a graph (where the edges weights can be defined taking into account the distance between the two connected nodes and the resources that each node can put on the communication) which is assumed to be connected. We will return to this in Section \ref{sec:spatialavg}. 

\begin{remark}
It should be noted that the assumed hypotheses about the temporal correlation structure of the noise and the source signal can be relaxed in several aspects. In first place, there is no need to assume that the spectrum of $s(t)$ is flat. Using again the formalism of Karhunen-Loève expansion the non-flat spectrum case can also be treated. Unfortunately, the final model is slightly more complex requiring a few more unknown parameters besides $\ve{c}$. Moreover, we can abandon the Gaussianity hypothesis of $s(t)$ if we include another unknown parameter related with the fourth-order moment of the expansion coefficients of $s(t)$. We have chosen to present the more restricted Gaussian model of $s(t)$ with flat spectrum over the system bandwidth to simplify presentation. As our main goal is to exploit the spatial correlation in the measurements in a distributed setting, we have assumed the simpler temporal correlation model for the noise and source signals presented above.
\label{remark:1}
\end{remark}

\section{GLR with local estimation} 
\label{sec:D-GLRT}
\subsection{Local estimate $\ve{c}_{\rm L-MLE}$}
\label{subsec:L-MLE}

The test in (\ref{eq:htpeq}) is basically a composite hypothesis testing problem. In particular, it is a parameter test \cite{Kay_SSP} (over $\ve{c}\succ \mathbf{0}_N$) because under both hypotheses the distribution is the same but with a different vector parameter: $\ve\theta=\ve 0_{N}$ under $\Hip_0$, and $\ve\theta=\ve c $ under $\Hip_1$. 
In order to perform the test, we have to build a statistic without using the unknown parameter $\ve c$, like the Wald or Rao test \cite{Kay_SSP,Levy_Det}, or to estimate it as in the generalized likelihood ratio (GLR) test. We follow the later approach, which has some asymptotic guarantees \cite{Kay_SSP,Levy_Det}.

The classical GLR statistic to test the hypotheses in (\ref{eq:htpeq}) is 
$T_G(\ve z)\equiv\frac{p(\ve z;\ve \theta = \hat{\ve c}_\text{G-MLE})}{p(\ve z; \ve \theta =\ve 0_N)}$, where $\hat{\ve c}_\text{G-MLE}$ is the (global) maximum likelihood estimator (MLE)\footnote{We call this estimator \emph{global} MLE to differentiate it from the local one  $\hat {\ve c}_{\rm{L-MLE}}$ to be defined next.} of $\ve c$ under $\Hip_1$ and $\ve z = \{\ve z_1,\dots,\ve z_L\}$.  We see from (\ref{eq:ck}) that each entry of the vector $\ve{c}$ is physically related with: i) the wireless link characteristics between the source and the corresponding sensor node, and ii) the second order statistical moment of the source. Also, from (\ref{eq:meanCoveq}), there is correlation between the energy measurements taken at different nodes, given by the term $\ve{c}\ve{c}^T$. 

It is also observed that the MLE for this model is difficult to compute even in a centralized scenario (where all measurements can be conveyed to a FC) given that the estimated value of each entry of $\ve{c}$ is a function of the energy measurements in all sensor nodes, that is, $\hat{c}_{{\rm G-MLE},k}\equiv f_k(\ve \{\ve z_l\}_{l=1}^{L})$ for each $k\in[1,N]$. In a distributed setting this would imply that each individual energy measurement in each sensor $k$ and time slot $l$ should be made available to all nodes in the network which is clearly not a practical solution. Even in the hypothetical case that all sensor measurements could be conveyed to each node across the network which would allow the computation of the MLE at each node, we would have the additional difficulty that not closed form mathematical solution for the MLE is available. Although each node could perform a numerical procedure to find the MLE, this would have a large computational load (that scales with network size $N$) which could impose a serious practical constraint, specially for nodes with limited computational capacity. 

Looking for a simpler approach to the computation of the MLE and taking into account that the local energy samples at each node should be sufficiently informative about the corresponding true value of $c_k$ for $k\in[1:N]$, we consider a \emph{local} MLE in sensor node $k$ which only use the locally sensed values $\left\{z_k(l)\right\}_{l=1}^L$. Therefore, we obtain the estimate $\hat {\ve c}_{\rm{L-MLE}}$ of $\ve c$ using the local estimates $\hat{c}_{{\rm L-MLE},k}\equiv g_k(\left\{z_k(l)\right\}_{l=1}^L)$, $k\in[1:N]$. In more precise terms, $\hat{c}_{{\rm L-MLE},k}$, is estimated using the model (\ref{eq:htpeq}) under $\Hip_1$ where sensor $k$ has access to its $L$ measurements whose distribution under $\Hip_1$ is $\{z_k(l)\}_{l=1}^{L} \overset{iid}{\sim} \N(\sqrt{M} c_k,(c_k+1)^2)$, that is the marginal distribution from $\N(\ve \mu(\mathbf{c}),\mathbf\Sigma(\mathbf{c}))$ for measurements at sensor node $k$. It is shown in the Appendix \ref{app:MLEloc} that $\hat{c}_{{\rm L-MLE},k}$ is given by:
\begin{multline}
\!\!\!\!\!\hat{c}_{{\rm L-MLE},k}\!=\!\frac{1}{2}\!\left(\sqrt{ (M\!+\!2\!+\!\sqrt{M} m_{k})^2\! +\! 4(p_z\!+\!\sqrt{M}m_k-1)}\right.\\
\left. - (M+2+\sqrt{M} m_{k})\right) \ k\in[1:N],
\label{eq:mlelocal}
\end{multline}
where $m_{k} = \textstyle\frac{1}{L}\sum_{l=1}^{L} z_{k}(l)$ and $p_{k}=\textstyle\frac{1}{L}\sum_{l=1}^{L} z_{k}^2(l)$. It is not difficult to see that the above computed local MLE $\hat {\ve c}_{\rm{L-MLE}}$ is the MLE for a signal model given by  $\{\ve z_l\}_{l=1}^{L} {\sim}  \prod_{l=1}^L\N(\ve\mu(\mathbf{c}),\text{diag}(\Sigma(\ve c)))$. That is, a model in which the correlation between the measurements at different sensor nodes is neglected. However, it is important to notice that, although we are neglecting this correlation, $\hat {\ve c}_{\rm{L-MLE}}$ is still an asymptotically consistent estimator of the true parameter $\mathbf{c}$. More importantly, it does not introduce a penalty in the asymptotic performance of the corresponding GLR statistic. We will show this in the next section.

\subsection{Asymptotic performance}
\label{sec:asymp}
In this section, we consider the asymptotic distribution, when $L\rightarrow\infty$, of the so called local GLR statistic in which we use the local MLE $\hat {\ve c}_{\rm{L-MLE}}$. We also provide a comparison with the full GLR statistic $T_G$ which uses the true global MLE $\hat{\ve c}_\text{G-MLE}$.  We prove that the local GLR statistic has exactly the \emph{same} asymptotic distribution. This strongly motivates the use of the local MLE $\hat {\ve c}_{\rm{L-MLE}}$ which is clearly easier to compute in a distributed setting.


\subsubsection{Full GLR}
It is well known that the distribution of the global GLR statistic $T_G$ is given by: \cite{Kay_SSP}:
\begin{equation}
2\log T_G(\ve z)\overset{a}{\sim}\left\{\begin{array}{ll}
\chi^2_N & \text{under } \Hip_0\\
\chi'^2_N(\lambda_g) & \text{under } \Hip_1,
\end{array}\right.
\label{eq:gglr}
\end{equation}
where the symbol $\overset{a}{\sim}$ means ``asymptotically distributed as when $L$ tends to infinity", 
$\chi^2_N$ is the chi-square distribution with $N$ degrees of freedom and $\chi'^2_N(\lambda_g)$ is the non-central chi-square distribution with $N$ degrees of freedom and non-centrality parameter $\lambda_g = L\ve c^T \ve i(\ve 0)\ve c$, where $\ve i(\ve 0)$ is the Fisher information matrix \cite{Kay_SSP_ET} evaluated at $\ve \theta=\ve 0$. As shown in Appendix \ref{app:MLEglob},  $\lambda_g = L(M+2)\|\ve c\|^2$.

\subsubsection{GLR with local MLE $\hat {\ve c}_{\rm{L-MLE}}$}
 We will obtain the asymptotic distribution of the local MLE and the local GLR statistic $T_{L}(\ve z)$ defined by 
\[T_{L}(\ve z) \equiv \frac{p(\ve z;\ve \theta = \hat {\ve c}_{\rm{L-MLE}})}{p(\ve z; \ve \theta =\ve 0)}.\]

We first note the particular structure of the joint pdf  $p(\ve z_l;\ve \theta)$ with $l\in[1:L]$ under $\mathcal{H}_1$ as shown in (\ref{eq:htpeq}) and (\ref{eq:meanCoveq}). From these equations and the discussion in the preceding section, it is clear that the marginalization of $p(\ve z_l;\ve \theta)$ over all components but the $k$-th, only depends on the $k$-th component of $\ve\theta$, $\theta_k$, and not on the whole vector $\ve \theta$, that is: 
\begin{align}
\int\cdots\int p(\ve z_l;\ve \theta)dz_1(l)\dots dz_{k-1}(l)dz_{k+1}(l)\dots dz_{N}(l) \nonumber\\ = p_k(z_k(l);\theta_k).\nonumber
\end{align}
However, the components of $\ve z_l$ are not independent, i.e., $p(\ve z_l;\ve\theta)\neq \prod_{k=1}^{N}p_k(z_k(l);\theta_k)$. Let $\mathcal{A}$ be a set of feasible points of $\ve \theta$ that includes the true parameter $\ve c$.  Define $\hat{\ve\theta}_{loc}^L\equiv\hat{\ve c}_{\rm{L-MLE}}$ and
consider the local MLE estimate $\hat{\ve\theta}_{loc}^L=[\hat{\theta}_{loc,1}^L,\dots,\hat{\theta}_{loc,N}^L]$. It is easy to see that its components are computed solving the following problem assuming that each sensor only has access to its own measurements:
\[\hat{\theta}_{loc,k}^L = \arg\max_{\theta_k\in \mathcal{A}^k} \frac{1}{L}\sum_{l=1}^{L}\log p_k(z_k(l);\theta_k),\] where $\mathcal{A}^k$ is the the projection of the set $\mathcal{A}$ on the $k$-th component. The following lemma states the asymptotic distribution of $\hat{\ve\theta}_{loc}^L$ and  the local GLR statistic $T_{L}$. The proof is presented in Appendix \ref{app:lem1}.
\begin{lemma}
Consider the following assumptions:
\begin{enumerate}
	\item[A1.] The first and second-order derivatives of the log-likelihood function are well defined and continuous functions.
	\item[A2.] $\Ex[\partial\log p_k(z_k(l);\theta_k)/\partial\theta_k]=0$, $k\in[1:N]$, $\forall l$.
	\item[A3.] The signal is \emph{weak}, i.e., $\|\ve c\|\!\leq\! c_0/\sqrt{L}$ for a constant $c_0$.
\end{enumerate}
We have that the asymptotic distribution of $T_L$ is:
\begin{align}
2\log T_{L}(\ve z) & \overset{a}{\sim}\left\{\begin{array}{ll}
\chi^2_N & \text{under } \Hip_0\\
\chi'^2_N(\lambda_{loc}) & \text{under } \Hip_1,
\end{array}\right.
\label{eq:lglr}
\end{align}
where $\lambda_{loc}=L(M+2)\|\ve c\|^2$ is the non-centrality parameter of the non-central chi-square distribution. 
\label{lemma:lemma_mis}
\end{lemma}
Notice that this result shows us that the distribution of the global (\ref{eq:gglr}) and the local (\ref{eq:lglr}) GLR statistics are asymptotically equal and, therefore, the asymptotic performance of both statistics is the same. This implies that our distributed hypothesis testing procedure has not penalties with respect to the centralized procedure, at least asymptotically. 
\begin{remark}
	Notice that the analysis of the asymptotic behavior of $\hat{\ve\theta}_{loc}^L\equiv\hat{\ve c}_{\rm{L-MLE}}$ is an instance of the \emph{mismatched} ML asymptotic performance problem \cite{White_1982}, \cite{Fortunati_Gini_Greco_Richmond_2017}. In our case, $\hat{\ve\theta}_{loc}^L$ is the so-called local MLE estimate based on the model $\prod_{k=1}^{N}p_k(z_k(l);\theta_k)$, which is clearly different from the \emph{true} data model $p(\ve z_l;\ve\theta)$. Our main interest is not in the asymptotic behavior of $\hat{\ve\theta}_{loc}^L$ but in the asymptotic behavior of the GLR statistic using this estimate. It is in this respect that we point out the relevance of Lemma \ref{lemma:lemma_mis}.  
\end{remark}


\section{Distributed computation of the testing statistic}
\label{subsec:diffusion}
In this section, we will consider the problem of adapting the GLR statistic that uses the local MLE $\hat {\ve c}_{\rm{L-MLE}}$ (denoted also as $\hat{\ve {c}}$ to keep notation uncluttered) in order to be efficiently computed and distributed across the network. 
From (\ref{eq:htpeq}), the GLR statistic can be written as:
\begin{align}
& \log T_{L}(\ve z) \nonumber\\
&= -\frac{L}{2}\log\det(\hat{\ve\Sigma}) \!+ \!\frac{1}{2} \sum_{l=1}^{L}\left\{\|\ve z_l\|^2 \!-\!\|\hat{\ve\Sigma}^{-\frac{1}{2}}(\ve z_l\!-\!\hat{\ve\mu})\|^2\right\}\nonumber\\
&\overset{(a)}{=} -\frac{L}{2}(\log(1+\bar{c}_1)+\bar{c}_2)\nonumber\\
&+\frac{1}{2}\sum_{l=1}^{L}\sum_{k=1}^{N}\left\{z_{k}^2(l)- \frac{(z_{k}(l)-\sqrt{M}\hat{c}_k)^2}{1+2\hat{c}_k}\right\}\nonumber\\
&\!+\! \frac{1}{2(1+\bar{c}_1)}\sum_{l=1}^{L}\left\{\sum_{k=1}^{N} (z_{k}(l)-\sqrt{M}\hat{c}_k)\frac{\hat{c}_k}{1+2\hat{c}_k}\right\}^2 
\label{eq:glr}
\end{align}
where both $\hat{\ve\mu}$ and $\hat{\ve\Sigma}$ are built using $\hat {\ve c}_{\rm{L-MLE}}$ given in (\ref{eq:mlelocal}). In the $(a)$-step, we used the Woodbury matrix inversion formula and the matrix determinant lemma to compute the closed forms of $\hat{\ve\Sigma}^{-1}$ and $\det(\hat{\ve\Sigma})$, respectively. We also defined
$\bar{c}_1\equiv\sum_{k=1}^N \frac{\hat{c}_k^2}{1+2 \hat{c}_k}$, $\bar{c}_2\equiv\sum_{k=1}^N\log(1+2 \hat{c}_k)$.

In first place, notice that the computation of $\bar{c}_1$ and $\bar{c}_2$ requires the spatial sum (through the index $k$) over the sensors of the quantities $\frac{\hat{c}_k^2}{1+2 \hat{c}_k}$ and $\log(1+2 \hat{c}_k)$ (which can be computed at each sensor using the local MLE $\hat {\ve c}_{\rm{L-MLE}}$). This spatial sum (proportional to the spatial averaging of the same quantities) can be computed with algorithms already developed in the literature \cite{xiao2004fast,sayed2013diffusion,cattivelli2010diffusion} or some of their variants, where the spatial average of a quantity is obtained by propagating  local neighborhood averages computed at each node. We formalize this in the next section.    

The second term in (\ref{eq:glr}) can also be written as a spatial sum of terms that can be computed locally at each sensor node. To show this, define for each $k\in[1:N]$:
\begin{equation}
u_k \equiv \sum_{l=1}^{L}\left\{z_{k}^2(l)- \frac{(z_{k}(l)-\sqrt{M}\hat{c}_k)^2}{1+2\hat{c}_k}\right\}.
\label{eq:u_k}
\end{equation}
As these terms can be computed locally at each sensor, the second term in (\ref{eq:glr}) can be written as $\frac{1}{2}\sum_{k=1}^Nu_k$, and again we can resort to a spatial averaging algorithm for its computation.

The third term in (\ref{eq:glr}), however is more complicated. As we can see, the summation in time and space do not commute in general. This implies that if we want to exactly compute this term we need to implement $L$ runs of the spatial averaging algorithm. When $L$ is large this could be inefficient in terms of energy, delay and bandwidth. Therefore, in order to simplify the computation of the distributed algorithm we do commute the summations and replace 
$
\frac{1}{L}\sum_{l=1}^{L}\left\{\sum_{k=1}^{N} (z_{kl}-\sqrt{M}\hat{c}_k)\frac{\hat{c}_k}{1+2\hat{c}_k} \right\}^2
$
by
\begin{align}
\left\{\sum_{k=1}^{N} \frac{1}{L}\sum_{l=1}^{L} (z_{kl}-\sqrt{M}\hat{c}_k)\frac{\hat{c}_k}{1+2\hat{c}_k} \right\}^2
=\nonumber \\
\left\{\sum_{k=1}^{N} (m_{k}-\sqrt{M}\hat{c}_k)\frac{\hat{c}_k}{1+2\hat{c}_k}\right\}^2,
\label{eq:approx}
\end{align}
 where $m_k$ was defined in Section \ref{subsec:L-MLE} and where the term inside the square can be computed as an spatial sum of terms that can be locally computed at each sensor. We will see in Section \ref{sec:numerical_results} that this replacement does not introduce a severe penalty in the algorithm performance in a wide range of signal-to-noise ratios. Defining for each $k\in[1:N]$:
$w_k \equiv (m_{k}-\sqrt{M}\hat{c}_k)\frac{\hat{c}_k}{1+2\hat{c}_k},$
$\bar{u}\equiv\sum_{k=1}^{N}u_k$ and $\bar{w} \equiv \sum_{k=1}^{N}w_k$, 
we can write the new statistic (which we call the \emph{fully distributed} statistic) as:
\begin{align}
\! T_{L-FD}(\ve z) &= -\frac{L}{2}(\log(1+\bar{c}_1)+\bar{c}_2)+\frac{\bar{u}}{2}
+ \frac{L \bar{w}^2}{2(1+\bar{c}_1)}.
\label{eq:FDstat}
\end{align}

\subsection{Spatial averaging algorithm}
\label{sec:spatialavg}
The above distributed statistic requires the computation of quantities $\bar{c}_1$, $\bar{c}_2$, $\bar{u}$ and $\bar{w}$ which as explained above are spatial sums over the different sensors in the network. Next we will generically refer to the sum $\bar{a}\equiv\sum_{k=1}^N a_k$, which will represent $\bar{c}_1,\bar{c}_2,\bar{u}$ or $\bar{w}$ accordingly. Each sensor node accesses to only a scalar value  $a_k\in \R$, $k\in\mathcal{N}\equiv[1:N]$ and it is desired to compute the average $\tilde{a} = \frac{1}{N}\sum_{k=1}^N a_k$ (or the sum $\bar{a} = N\tilde{a}$) at each node in a distributed manner and with minimal resources allocated to the exchanges between the nodes.
 
Assuming that the nodes only communicates with their neighbors through error-free channels, the spatial averages can be computed via a distributed diffusive procedure such as in \cite{xiao2004fast,sayed2013diffusion,cattivelli2010diffusion}.
Between all the existing possibilities, we will consider a simple but effective algorithm usually called \emph{local-degree weights} distributed averaging algorithm \cite{xiao2004fast}. In more precise terms, consider a network (modeled as a connected graph) $\mathcal{G}=(\mathcal{N},\mathcal{E})$ consisting of a set of nodes $\mathcal{N}$ and a set of edges $\mathcal{E}$, where each edge $\{i,j\}\in\mathcal{E}$ is an unordered pair of distinct nodes. The set of neighbors of node $i$ is denoted by $\mathcal{N}_i = \{j\in\mathcal{N}|\{i,j\}\in \mathcal{E}\}$. Notice that the sensor $i\notin\N_i$. See Fig. \ref{fig:sensingnode} (a) for a graphical description.

The average value $\tilde{a}$ can be computed iteratively as:
\begin{equation}
a_k(t) = W_{kk} a_k(t-1) + \sum_{j\in\mathcal{N}_k} W_{kj} a_j(t-1),\ \ k\in\N, \ t\in\mathbb{N}
\label{eq:spatialAve}
\end{equation}
where $a_k(t)$ is the average after $t$ iterations (or message exchanges between the nodes), $a_k(0)=a_k$ is the initial value and $W_{kj}$ is the weight on $a_j(t-1)$ at the node $k$. These set of equations can be succinctly written using using a matrix formulation. To this purpose, let $\ve a(t) \equiv [a_1(t),\dots,a_N(t)]^T$. Then, the iterative equation in its matrix form is:
$$\ve a(t)=\mathbf W \ve a(t-1) = \mathbf W^t \ve a(0),\ t= 1,2,\dots$$
where $\mathbf W$ is the matrix of weights with elements $W_{kj}=(\mathbf W)_{kj}$. Considering local communication only, i.e., each node broadcasts its local value at iteration $t$ only to the nodes in its neighborhood, we have that for each $k\in\N$, $W_{kj}=0$ for $j\notin\N_k$ and $j\neq k$. Thus, the feasible weight matrices must satisfy a sparsity pattern given by the network connectivity: $\mathbf W\in \mathcal{S}$, where $\mathcal{S} = \{\mathbf W\in \R^{N\times N} |W_{kj} = 0 \text{ if } \{k,j\} \notin \mathcal{E}\text{ and } k\neq j\}$. 

The optimum weights matrix $\mathbf W$ in terms of the asymptotic convergence factor can be computed by solving  semidefinite program (SDP), assuming that $\mathbf W$ is a symmetric matrix \cite{xiao2004fast}. Although this procedure guaranties the fastest convergence of $\ve a(t)$ to the average vector $\tilde{a} \ve 1_N$ when $t\rightarrow\infty$, each node should be aware of its corresponding weights to perform the average. This would require to solve the mentioned SDP program in a distributed manner or, optionally, in a centralized manner and then to communicate the corresponding weights to each node. In both cases, the complexity of this procedure is high. 

A simpler way is to select the weights directly, without any optimization procedure. This, for example, can be achieved with local-degree weights \cite{xiao2004fast} where the convergence to the required average is guaranteed given that graph is not bipartite. Although with this choice we are sacrificing speed of convergence to the desired average, we will use it because it requires a minimal knowledge, at each node, about the network topology. This is certainly a very much desired feature, specially for large and/or rapidly changing networks. 

The weights are defined as follows. Assume arbitrarily a direction for each edge of the graph. Let $P\equiv|\mathcal{E}|$ be amount of edges of the graph, and define the incident matrix $\mathbf A\in\R^{N\times P}$ as\footnote{It can be proved that neither the election of the edges direction nor the sign of the ones in the incident matrix modify the weights defined in (\ref{eq:weights}).}
$$ A_{kj} = (\mathbf A)_{kj}= \left\{\begin{array}{rl}
1, & \text{if edge $j$ starts from node $k$,}\\
-1,& \text{if edge $j$ ends at node $k$,}\\
0, & \text{otherwise.}
\end{array}\right.$$ 
Considering symmetric weights, each edge is associated with a unique weight $w_l=W_{kj}=W_{jk}=1/\max(d_k,d_j)$, where edge $l\in[1:P]$ connects nodes $k$ and $j$ and $d_k$ is the degree of node $k$, i.e., the number of neighbors of node $k$. Letting $\ve w\in \R^P$ with components $(\ve w)_l=w_l$, the matrix of weights can be written as 
\begin{equation}
\mathbf W = \mathbf{I}_N-\mathbf A \diag(\ve w) \mathbf A^T.
\label{eq:weights} 
\end{equation}
It should be easy to notice that, as explained above, this construction of the weights depends only of local information about network connectivity at each node (only the degree of each node is required). Clearly, in a WSN, this information is available in each node at the network layer of the communication stack.

Algorithm \ref{fd} summarizes the steps required to compute the fully distributed statistic (\ref{eq:FDstat}). Several stopping criteria can be considered in the iterative computation of the spatial average (\ref{eq:spatialAve}). For example, we can consider stopping criteria as a fixed number of exchanges, or a fixed number of exchanges after no significant changes in each $a_k(t)$, $k\in[1:N]$, is observed. Although this a important aspect of the distributed calculation procedure for the spatial averaging,
in this work, we will consider the former option to evaluate numerically the algorithm performance in Section \ref{sec:numerical_results}. 

It is important to consider the total number of messages that the nodes need to exchange in order to compute $T_{L-FD,k}$, for each $k\in [1:N]$. Let $N_{it}$ be the predefined number of message exchanges or iterations established for computing the fully distributed statistic. As already mentioned, it needs to compute 4 spatial sums: $\bar{c}_{1,k}$, $\bar{c}_{2,k}$, $\bar{u}_k$ and $\bar{w}_k$, and for each of them, $N\times N_{it}$ message transmissions are needed, given that each node broadcasts its data to its neighborhood. Then, a total of  $4 N N_{it}$ transmissions are needed. On the other hand, the local statistic $T_L$ in (\ref{eq:glr}) requires to compute $\bar{c}_{1,k}$, $\bar{c}_{2,k}$, $\bar{u}_k$ (as in $T_{L-FD}$) and $L\times N\times N_{it}$ broadcast transmissions to compute the last term in (\ref{eq:glr}). It makes a total of $(3+L)N N_{it}$ transmissions, which is typically much greater that $ 4 N N_{it}$ when the time slots for energy computation at each node satisfy $L\gg 1$. This analysis shows the advantage of $T_{L-FD}$ over $T_L$, in terms of communication resources, for the source detection problem in a distributed scenario.   

It is important to remember how the use of the local MLE $\hat {\ve c}_{\rm{L-MLE}}$ allowed us to have a fully and efficient distributed statistic $T_{L-FD}(\ve z)$ in terms of communication exchanges over the network. It should be also clear, that this would have been impossible with the global MLE solution $ \hat{\ve{c}}_{\rm G-MLE}$. At this point we would ask ourselves if the use of the local MLE, although important from a practical point of view in the distributed setting, will bring some performance penalization of the hypothesis testing problem, in the non-asymptotic regime, with respect to the case in which the global MLE solution is employed. This will be analyzed in the following section via numerical simulations.

\begin{algorithm}[t]
\caption{Computing of $T_{L-FD}$}\label{fd}
\begin{algorithmic}[1]
\State $\triangleright$ Distributed computation of the fully distributed statistic
\For {$k=1,\dots,N$}{ (\it simultaneously at each sensor)}
\State Compute the local estimate $\hat{c}_k$ using eq. (\ref{eq:mlelocal}).
\State $\bar{c}_{1,k}=$ \Call{SpatialSum}{$\frac{\hat{c}^2_k}{1+2\hat{c}_k}$} \Comment{Compute the sum through the nodes}
\State $\bar{c}_{2,k}=$ \Call{SpatialSum}{$\log{(1+2\hat{c}_k)}$}
\State $\bar{u}_{k}=$ \Call{SpatialSum}{$u_k$}
\State $\bar{w}_{k}=$ \Call{SpatialSum}{$w_k$}
\State $T_{L-FD,k}=$ \Call{ComputeStat}{$\bar{c}_{1,k},\bar{c}_{2,k},\bar{u}_{k},\bar{w}_{k}$}
\If {$T_{L-FD,k}<\gamma$} {Sensor $k$ decides $\Hip_0$,} \Comment{$\gamma$ is the predefined threshold of the test.}
\Else { Sensor $k$ decides $\Hip_1$.} 
\EndIf  
\EndFor
\State$\triangleright$ Definition of functions
\Function{SpatialSum}{$a_k$}\Comment{Compute iteratively the spatial sum of ${a}_k$, $\bar{a}_k$.}
\State $a_k(0)=a_k$\Comment{Initial condition for $t=0$.}
\State $t = 0$
\While{Stop criterion not met}
\State $t = t+1$
\State Compute the spatial average $a_k(t)$ using (\ref{eq:spatialAve}) with weights      (\ref{eq:weights}).
\EndWhile
\State \textbf{return} $N a_k(t)$ \Comment{Return the sum $\bar{a}_k$}
\EndFunction

\Function{ComputeStat}{$\bar{c}_{1,k},\bar{c}_{2,k},\bar{u}_{k},\bar{w}_{k}$}
\State Compute $T_{L-FD,k}$ using (\ref{eq:FDstat}).
\State \textbf{return} $T_{L-FD,k}$ 
\EndFunction

\end{algorithmic}
\end{algorithm}

\section{Numerical results}
\label{sec:numerical_results}

In this section, we compare the performance of the proposed fully distributed statistic with a finite amount of samples per node against the asymptotic results. We also numerically evaluate the ability of the network to achieve consensus about the final decision (source present or absent) between its nodes.

\begin{figure}
	\centering
	\includegraphics[width=\linewidth]{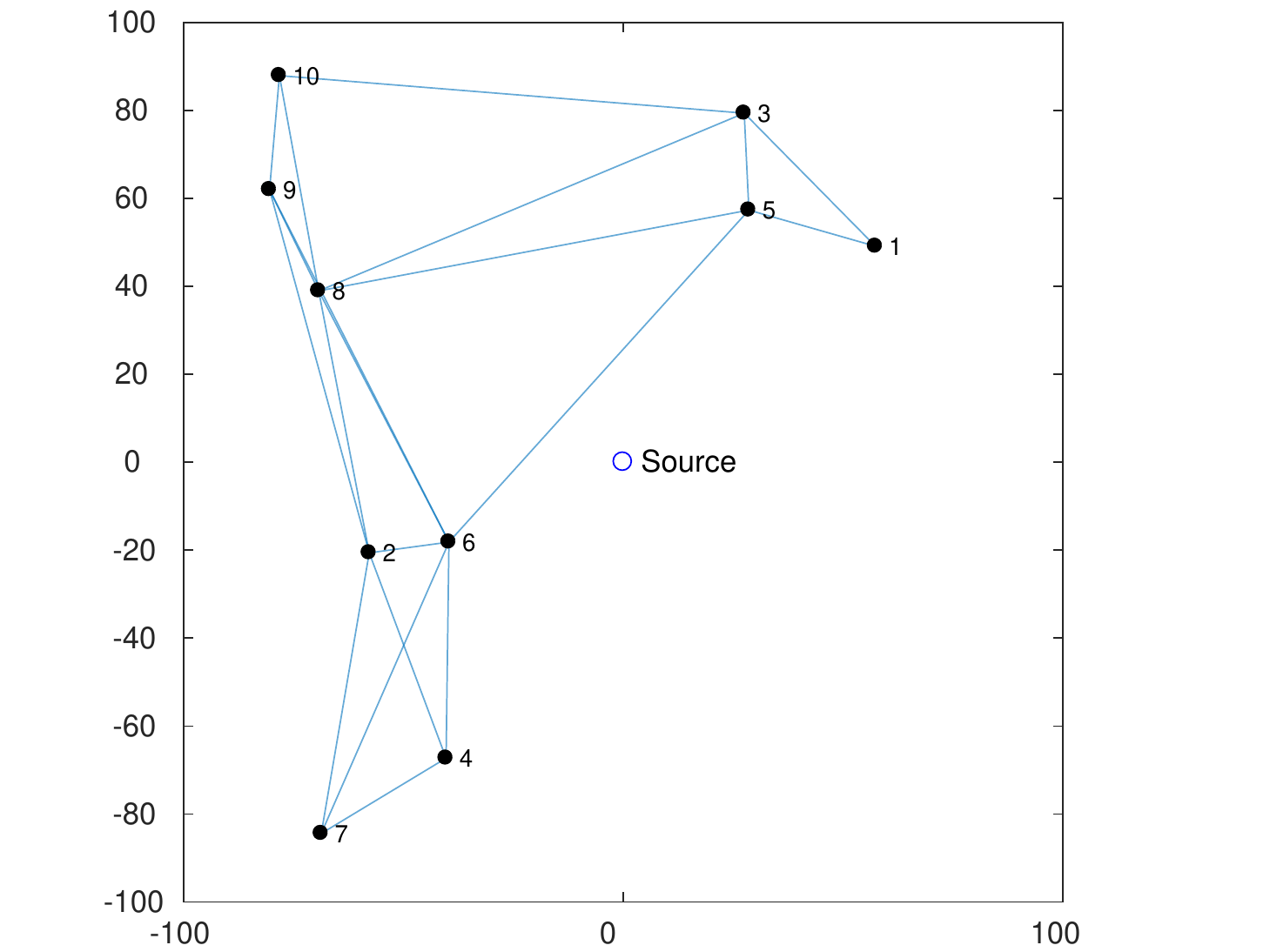}
	\caption{Randomly generated network with $N=10$ sensor nodes (black dots) and $|\mathcal{E}|=20$ edges connected with its neighbors.}
	\label{fig:network}
\end{figure}

We consider the network represented through its graph shown in the Fig. \ref{fig:network} with $N=10$ nodes and $|\mathcal{E}|=20$ edges, and the source located in the center $(0,0)$. This network was randomly generated following \cite{xiao2004fast}. First we randomly generated 10 nodes, uniformly distributed on a square of $200\times 200 \ \rm m^2 $. We impose that two nodes are connected by an edge if their distance is less than a predefined threshold. Then we increase the threshold until the total number of edges is $20$ and check that the resulting graph is connected. 

We set the following parameters which could be assumed for sensing, for example, a TV signal in the 400-800 MHz UHF band, where the bandwidth of each channel is 6 MHz \cite{cabric2004implementation}. Therefore, we take $W = 6$ MHz\footnote{A Cognitive Radio receiver with energy detector could be implemented as in \cite{cabric2004implementation} using a analog-to-digital converter followed by a N-FFT operation, an averaging and a square device. In this context, the frequency bin spacing is $W = 6$ MHz.}. The observation window is set to $\tau = 41.6\mu$s and is divided in $L=50$ time-slots of duration each $T=\frac{\tau}{L}=0.83\mu$s. Then, the time-bandwidth product $M=2WT=10$. Finally, we take the path loss to be $\alpha = 4$. The non-centrality parameter $\lambda=\lambda_{loc}=\lambda_{g}$ depends on the quotient $\sigma^2_s/\sigma^2_v$ through $\ve c$. Thus, when the source is present, $\sigma^2_s/\sigma^2_v$ is adjusted to achieve the value $\lambda$ shown (in dB) in each figure.

Before showing the numerical results, we define the miss-detection and the false alarm probability of a statistic $T$ for a predefined threshold $\gamma$ as $P_\text{md}\equiv\Pr(T<\gamma|\Hip_1)$ and $P_\text{fa}\equiv\Pr(T>\gamma|\Hip_0)$. The detection probability is $P_\text{d}=1-P_\text{md}$. 

\begin{figure}
\centering
\includegraphics[width=\linewidth]{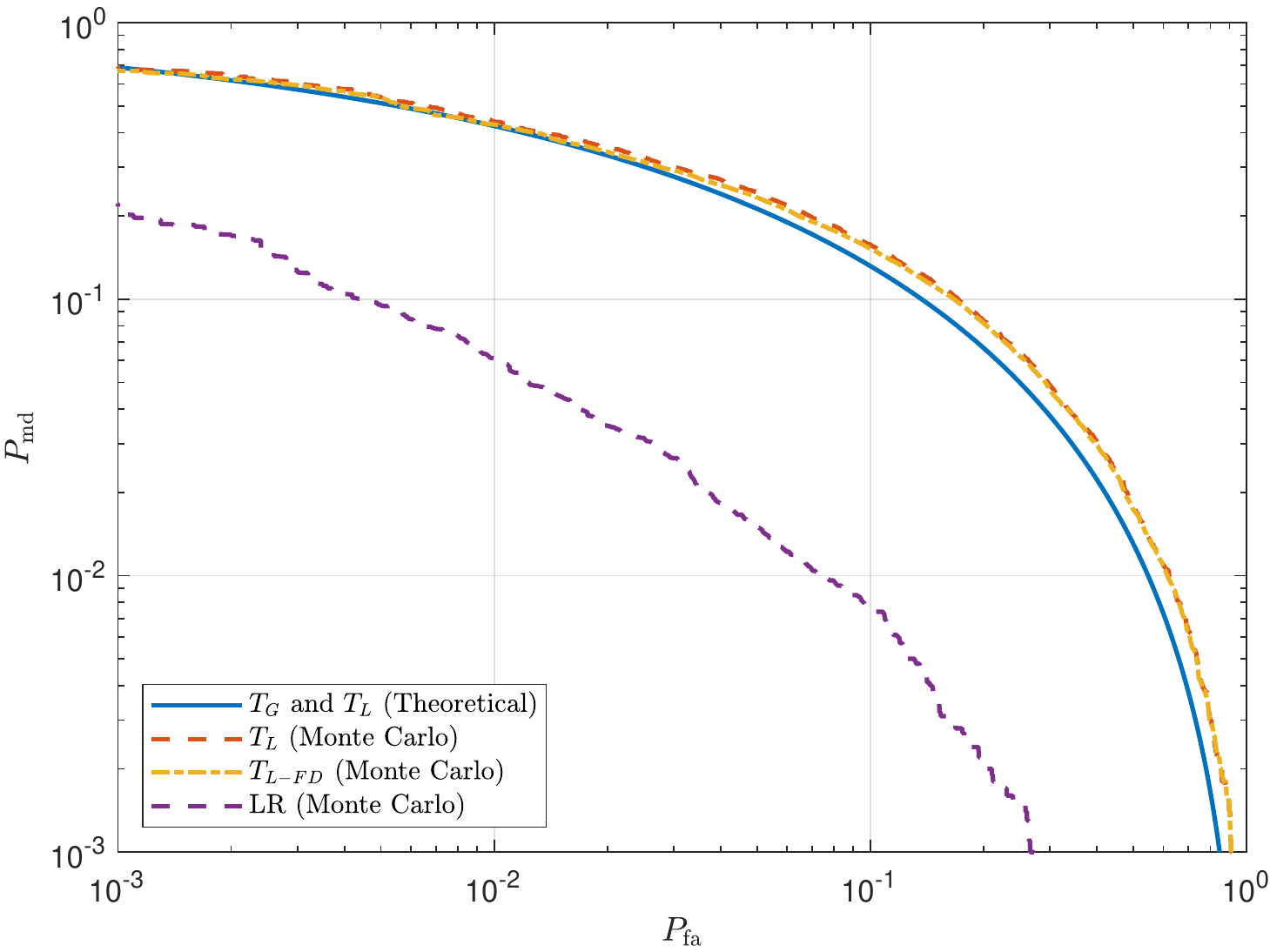}
\caption{CROC for $\lambda=12$ dB for the considered statistics.}
\label{fig:ROC}
\end{figure}

In Fig. \ref{fig:ROC}, we set $\lambda =12$ dB and plot several complementary receiver operating characteristics (CROC) for the presented statistics. First, we plot the (theoretical) asymptotic performance of the GLR test using global MLE estimation $\hat{\ve c}_{G-MLE}$, $T_G$, and local MLE estimation $\hat{\ve c}_{L-MLE}$, $T_L$ (c.f. (\ref{eq:gglr}) and (\ref{eq:lglr}), respectively). We also evaluate the performance of all statistics for a finite amount of measurements (i.e. $L=50$) generating $10^4$ Monte Carlo runs.
We see that the performance of the statistic $T_{L}$ matches very well with the theoretical asymptotic performance. We also see that the fully distributed statistic $T_{L-FD}$ with $N_{it}=20$ iterations (see Algorithm \ref{fd}) has a similar performance to $T_{L}$ (the curves are superposed). This shows that the replacement in (\ref{eq:approx}) works well, allowing for a fully distributed computation without introducing any significant loss in the performance. Finally, the behavior of the optimal likelihood ratio (LR) test, computed through the method of Monte Carlo, is also plotted only to have a purely theoretical reference. It is not possible to implement this test in practice, due to the fact that it requires the exact knowledge of the parameters under $\Hip_1$.

\begin{figure}
\centering
\includegraphics[width=\linewidth]{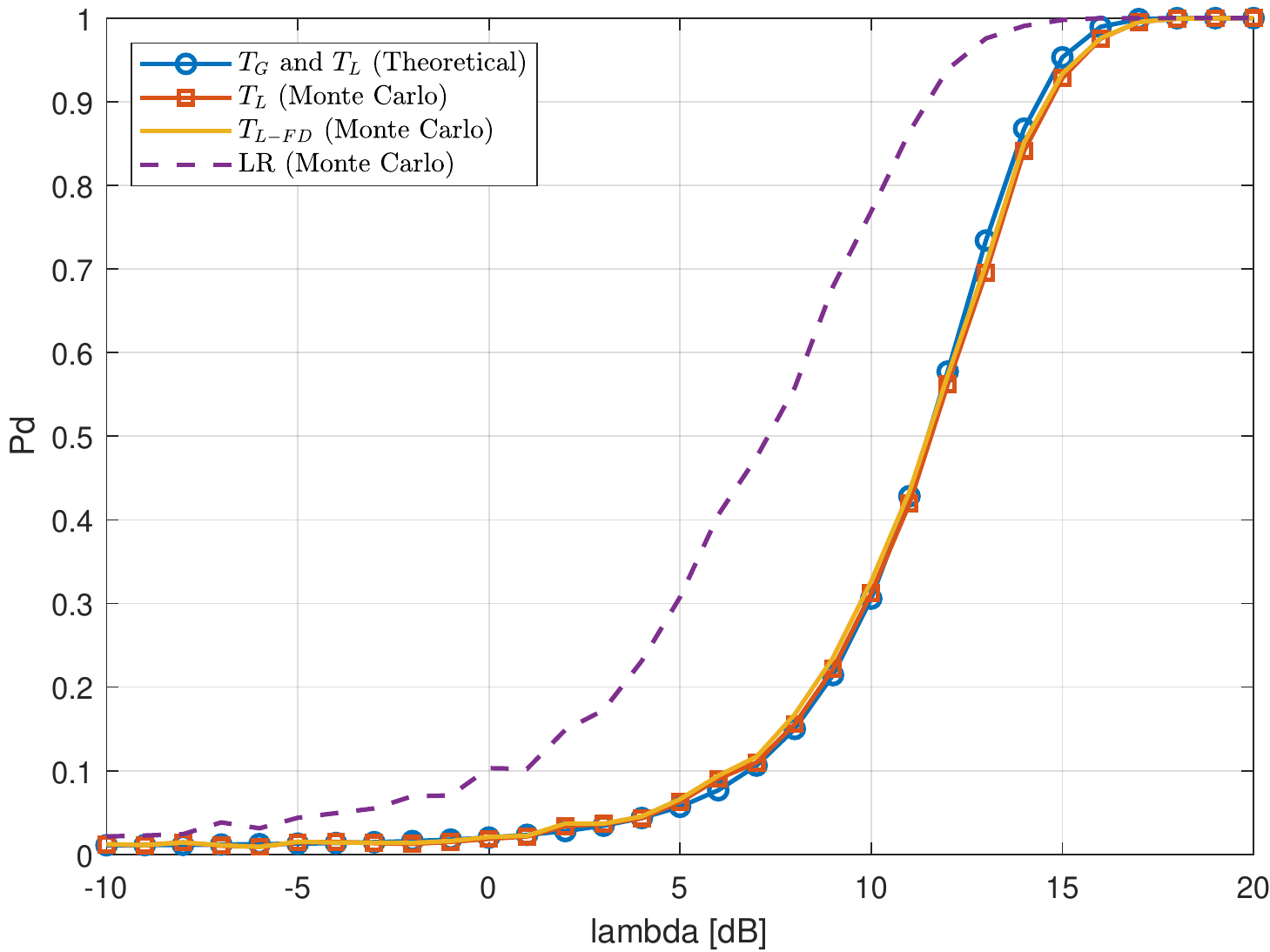}
\caption{Detection probability vs $\lambda$ for a fixed false alarm probability $P_\text{fa}=0.01$.}
\label{fig:PdVsSNR}
\end{figure}

In Fig. \ref{fig:PdVsSNR}, we plot the detection probability for a wide range of $\lambda$ for a fixed false alarm probability $P_\text{fa}=10^{-2}$. As in the previous figure, the curves match very well in all the range validating again the performance of $T_{L-FD}$. Again, $N_{it}=20$ iterations through the nodes are used to compute the fully distributed algorithm. We emphasize that the fully distributed algorithm has almost the same performance than the global GLR statistic and has a loss of about 3 dB in the parameter $\lambda$ for $P_\mathrm{d}=0.9$ with respect to the unrealizable likelihood ratio test. Notice also that the parameter $\lambda$ determines the performance of the tests and can be written as $\lambda = L(M+2)N \rho_\text{avg}$, where $\rho_\text{avg}\equiv\|\ve c\|^2/N$ can be interpreted as a signal-to-noise ratio averaged through the sensor nodes. In general, $\rho_\text{avg}$ depends on the coverage range of the network, the signal propagation model, the source power and the noise power, and cannot be chosen freely. However, the designer of the sensor network has freedom to select the remaining parameters: $L$ (number of sensing time slots), $M$ (time-bandwidth product) and $N$ (number of sensors). Thus, the analytical characterization of the network performance obtained in this work gives a good starting point for designing  wireless sensor networks for the application of source detection without the need of a FC and using simple and cheap energy detectors at each sensor node. 

\begin{figure}
\centering
\includegraphics[width=\linewidth]{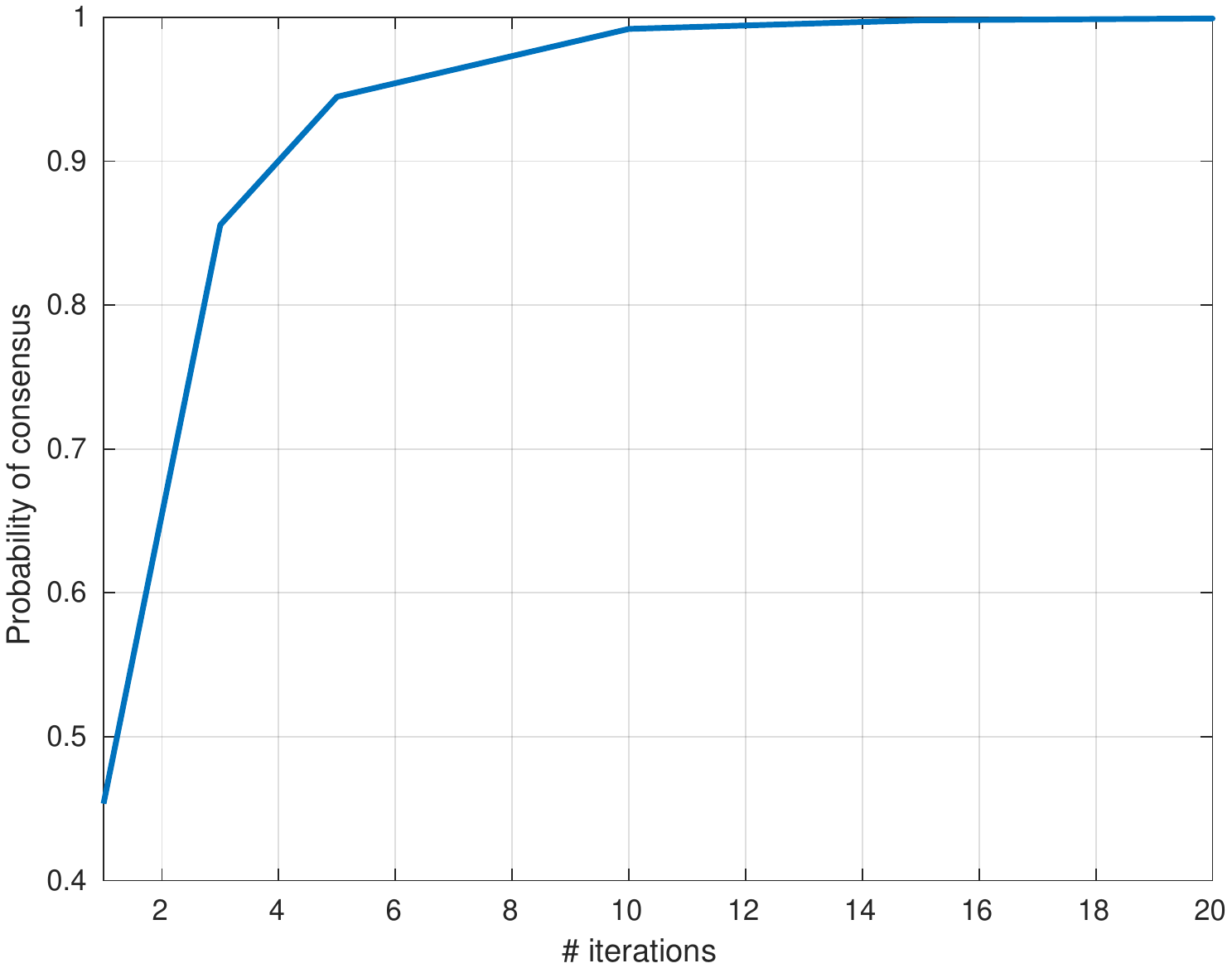}
\caption{Probability of all sensors make the same decision vs number of iterations for $\lambda = 17$ dB, $P_\text{fa}= 0.0048$ and $P_\text{md}= 0.0097$}
\label{fig:PconsensusVsIter}
\end{figure}
Finally, we evaluate the ability of the network to achieve consensus between its nodes using the fully distributed statistic. As each sensor computes, in principle and for a finite number of iterations (or communication exchanges with its neighbors), a different statistic, the decision at each node could be not the same. In order to quantify this, we define the probability of consensus of the network as the probability that \emph{all} sensors make the same decision about the presence or absence of the source. Obviously, this probability depends strongly on the graph of the network. In Fig. \ref{fig:PconsensusVsIter}, we plot this probability computed through the method of Monte Carlo against the number of iterations used to build the fully distributed statistic. It can be seen that for the network of Fig. \ref{fig:network}, only 4 iterations (or communication exchanges at each node) are needed to achieve a 90\% probability of consensus and that for 10 iterations the consensus is practically a certain event. This allows us to conclude that the spatial average procedure proposed is economical in terms of resources consumption.

\section{Conclusions}
In this paper we derived a signal model for nodes in a WSN that implement energy detectors to perform a test and decide if a source emitting a stochastic signal is present or not. We then built a cooperatively fully distributed algorithm suitable for a WSN and analyzed its performance. We showed that even though the measurements from different sensors are correlated, the parameters of the test can be estimated locally without asymptotically performance loss. This was analytically proved by computing the asymptotic distribution of the GLR test with the proposed local MLE estimator of the unknown parameters, showing that it is the same as the global GLR distribution which consider the global MLE estimation of the same quantities. This allows to quantify the performance of the proposed test and evaluate its dependence with the main parameters of the problem to design a WSN for this kind of applications. Finally, we showed that few iterations of the fully distributed algorithm are sufficient to achieve a high probability of consensus between the nodes, showing the usefulness of the proposed algorithm in a WSN with limited resources. 

\appendices
\section{Derivation of the local MLE}
\label{app:MLEloc}
Consider sensor $k$ and its energy measurements denoted by  $\ve z^k = [z_k(1),\dots,z_k(L)]^T$. The distribution of $\ve z^k$ under $\Hip_1$ is $\ve z^k \sim \N(\sqrt{M} c_k\ve 1_L,(c_k+1)^2 \mathbf{I}_L)$. The log-likelihood is given by (neglecting terms which do not depend on $c_k$):
\[\log p(\ve z^k; c_k) \propto -L\log(1+c_k)-\frac{1}{2}\left\|\frac{\ve z^k-\sqrt{M}c_k \ve 1_L}{1+c_k}\right\|^2.\]
Deriving and setting to zero it is easy to show that the optimal solution for the maximum of the log-likelihood has to satisfy the following quadratic equation:
\begin{multline}
c_k^2+\left(M+2+\sqrt{M} m_k \right)c_k 
-\left(p_k
+\sqrt{M}m_k -1 \right) = 0.\nonumber
\end{multline}
Keeping the positive root, we obtain the MLE estimator (\ref{eq:mlelocal}). 

\section{Computation of $\lambda_{g}$}
\label{app:MLEglob}
To evaluate $\lambda_{g}$ we need to compute the following Fisher information matrix evaluated at $\ve \theta=\ve 0$
\begin{equation}
\ve i(\ve 0)= \Ex_{\ve\theta =\ve c}\left.\left[\frac{\partial\log p(\ve z_l;\ve\theta)}{\partial\ve\theta} \frac{\partial\log p(\ve z_l;\ve\theta)}{\partial\ve\theta^T}\right]\right|_{\ve\theta=\mathbf{0}}
\label{eq:FisherGlobal}
\end{equation}
where the expectation is taken with respect to $p(\ve z_l;\ve \theta=\ve c)$.
When $p(\ve z_l;\ve \theta)$ with $l\in[1:L]$ is Gaussian, as in (\ref{eq:htpeq}), the $(i,j)$-th component of the Fisher information matrix can be computed as \cite{Kay_SSP_ET}
\begin{align}
[\ve i(\ve\theta)]_{kj} &=\frac{\partial\ve\mu(\ve\theta)}{\partial\theta_k}\mathbf\Sigma(\ve\theta)^{-1}\frac{\partial\ve\mu(\ve\theta)}{\partial\theta_j}\nonumber\\
&+\frac{1}{2}\tr\left[\mathbf\Sigma(\ve\theta)^{-1}\frac{\partial\mathbf\Sigma(\ve\theta)}{\partial\theta_k} \mathbf\Sigma(\ve\theta)^{-1} \frac{\partial\mathbf\Sigma(\ve\theta)}{\partial\theta_j}\right]. 
\end{align}
From (\ref{eq:meanCoveq}), we have $\frac{\partial\ve{\mu}(\ve \theta)}{\partial\theta_k} = \sqrt{M}\ve e_k$, and $\frac{\partial\mathbf\Sigma(\ve \theta)}{\partial\theta_k} = (\ve\theta \ve e_k^T + \ve e_k \ve\theta^T)+ 2\ve e_k\ve e_k^T$, where $\ve e_k\in\R^N$ is the canonical vector with value $1$ in the $k$-th coordinate and $0$ otherwise. Then, $\ve i(\ve 0)= (M+2)I_N$ and $\lambda_g = L(M+2)\|\ve c\|^2$.

\section{Proof of Lemma 1}
\label{app:lem1}
The following proof follows the spirit of that one found in \cite{Kay_SSP,Kay_SSP_ET}, although it has important modifications. In the following we will make use of the following result:
\begin{theorem} [Mean value theorem] \cite[Th. 12.9]{apostol1974mathematical}
	Let $B$ be an open subset of $\R^N$ and assume that $\ve f: B \rightarrow \R^N$ is differentiable at each point of $B$. Let $\ve x$ and $\ve y$ be two points in $B$ such that the segment $S(\ve x, \ve y)= \{t\ve x+ (1-t)\ve y: t\in [0,1] \} \in B$. Then for every vector $\ve a$ in $\R^N$ there is a point $\ve z\in S(\ve x, \ve y)$ such that
	\begin{equation}
	\ve a^T (\ve f(\ve y) - \ve f(\ve x)) = \ve a^T J(\ve w)(\ve y - \ve x),
	\end{equation}
	where $J(\ve w)$ is the Jacobian matrix of $\ve f$ evaluated in $\ve w$, i.e., $[J]_{i,j}=\frac{\partial f_i}{\partial x_j}$, where $\{f_i\}$ are the components of $\ve f$. It is important to remark that $\ve w$ depends on $\ve a$.
	\label{theo:mean}
\end{theorem}
\subsection{Asymptotic distribution of the local MLE}
\label{app:lem1-a}
Firstly, we prove the consistency of the local MLE. Consider $k\in[1:N]$. Given that $\{z_k(l)\}_{l=1}^L$ are i.i.d. under both $\Hip_0$ and $\Hip_1$, we can apply the weak law of large numbers (LLN):   
\begin{align}
\frac{1}{L}\sum_{l=1}^L \log p_k(z_k(l);\theta) \overset{p}{\underset{L\rightarrow\infty}{\longrightarrow}} &\Ex_{\theta_k= \theta_k^*}[\log p_k(z_k(l);\theta)] \label{eq:LLN1}\\
& \equiv \int \log p_k(u;\theta)p_k(u;\theta_k^*)du \label{eq:LLN2}
\end{align}
where the sum converges in probability and $\theta_k^*$ is the true parameter of the distribution ($\theta_k^* = 0$ under $\Hip_0$ and $\theta_k^* = c_k$ under $\Hip_1$).
Consider now the following inequality derived from the non-negative property of the Kullback-Leibler divergence \cite{Cover_Thomas_2006} between $p_k(u;\theta_1)$ and $p_k(u;\theta_2)$ for arbitrary $\theta_1$ and $\theta_2$: 
\[\int \log p_k(u;\theta_1)p_k(u;\theta_1)du\geq \int \log p_k(u;\theta_2)p_k(u;\theta_1)du. \]
Then, it is clear that (\ref{eq:LLN2}) is maximized for $\theta = \theta_k^*$, and by a suitable continuity argument \cite{Vaart_2000}, the LHS of (\ref{eq:LLN1}) must also be maximized for $\theta = \theta_k^*$. Therefore, $\hat{\theta}_{loc,k}^L\overset{p}{\rightarrow}\theta_k^*$ as $L\rightarrow\infty$ $\forall k = [1:N]$. In vector form we have $\hat{\ve\theta}_{loc}^L\overset{p}{\rightarrow}\ve\theta^*$ as $L\rightarrow\infty$. Thus, the local MLE estimator is consistent.

Now, we derive the asymptotic distribution of $\hat{\ve\theta}_{loc}^L$. 
Let 
$$\ve \psi(\ve z_l;\ve\theta) = \left[\frac{\partial\log p_1(z_1(l); \theta_1)}{\partial\theta_1},\dots,\frac{\partial\log p_N(z_N(l); \theta_N)}{\partial\theta_N}\right]^T.$$
By definition, the MLE must satisfy 
\begin{equation}
\left.\frac{1}{L}\sum_{l=1}^{L} \ve\psi(\ve z_l;\ve\theta) \right|_{\ve\theta=\hat{\ve\theta}_{loc}^L}=\ve 0.
\label{eq:MLE_deriv}
\end{equation}
Consider Theorem \ref{theo:mean} with $\ve f(\ve \theta) = \frac{1}{L}\sum_{l=1}^{L} \ve\psi(\ve z_l;\ve\theta)$, $\ve x = \ve\theta^*$ and $\ve y =\hat{\ve\theta}_{loc}^L$, then we have
\begin{align*}
\ve a^T \left( \frac{1}{L}\sum_{l=1}^{L} \ve\psi(\ve z_l;\hat{\ve\theta}^L_{loc})- \frac{1}{L}\sum_{l=1}^{L} \ve\psi(\ve z_l;\ve\theta^*)\right) \\
= \ve a^T J(\ve w^L)(\hat{\ve\theta}^L_{loc}-\ve\theta^*), \ \ \forall \ve a\in\mathbb{R}^N
\end{align*}  
where $\ve w^L$ belongs to the segment $S(\hat{\ve\theta}^L_{loc},\ve\theta^*)$ and depends on $\ve a$. 
Assuming that the Jacobian matrix of $\ve f(\ve \theta)$, $J(\ve w^L)$ is invertible and using (\ref{eq:MLE_deriv}) in the previous equation,
\begin{align}
\ve a^T \left\{\frac{1}{\sqrt{L}}J(\ve w^L)\left( J(\ve w^L)^{-1} \frac{1}{\sqrt{L}}\textstyle\sum_{l=1}^{L} \ve\psi(\ve z_l;\ve\theta^*)\right.\right. \nonumber\\
\left.\left.+ \sqrt{L}(\hat{\ve\theta}^L_{loc}-\ve\theta^*)\right)\right\}=0, \ \ \forall \ve a.
\label{eq:mvt}
\end{align}  
By consistency of the estimator $\hat{\ve\theta}^L_{loc}$, the segment $S(\hat{\ve\theta}^L_{loc},\ve\theta^*)$ becomes the point $\ve\theta^*$ and 
$\ve w^L\overset{p}{\rightarrow}\ve\theta^*$ as $L\rightarrow\infty$. 
Thus, the expression inside the parenthesis in (\ref{eq:mvt}) becomes independent of $\ve a$ as $L\rightarrow\infty$, and therefore, it must converge in probability to $\ve 0$. Then,
using the continuity of the second-order partial derivatives of the log-likelihood function to apply the Continuous Mapping Theorem (CMT)\cite{Vaart_2000}, we have $J(\ve w^L)\overset{p}{\rightarrow}J(\ve\theta^*)$ where 
\begin{equation*}
[J(\ve{\theta}^*)]_{ij}= \left.\Ex_{\ve{\theta}=\ve{\theta}^*}\left[\frac{\partial^2\log p_i(z_i(l); \theta_i)}{\partial\theta_i\partial\theta_j}\right]\right|_{\theta_i=\theta_i^*}. 
\end{equation*} 
Clearly, $J(\ve{\theta}^*)$ is a diagonal matrix given the fact that $p_i(z_i(l); \theta_i)$ is a function only on $\theta_i$. 
Additionally, by the central limit theorem (CLT)
\begin{equation*}
\frac{1}{\sqrt{L}}\sum_{l=1}^{L} \ve\psi(\ve z_l;\ve\theta^*) \overset{a}{\sim} \N(\ve 0,\tilde{\ve i}(\ve \theta^*)),
\end{equation*}
where the mean of the Gaussian distribution is $\ve 0$ by assumption A2 and its covariance matrix is defined by the \emph{local} Fisher information matrix, given its resemblance to the Fisher information matrix (\ref{eq:FisherGlobal})\footnote{Notice that the difference between the Fisher information matrix and the local one is the probability distribution inside the expectation: the full joint distribution appears in the former, while the marginal distribution corresponding to each node are considered in the local case.}
\begin{align}
& [\tilde{\ve i}(\ve \theta^*)]_{ij} =\nonumber\\ 
& \left.\Ex_{\ve\theta=\ve \theta^*}\left[\frac{\partial\log p_i(z_i(l);\theta_i)} {\partial\theta_i} \frac{\partial\log p_j(z_j(l);\theta_j)}{\partial\theta_j}\right]\right|_{\ve\theta=\ve\theta^*}.
\label{eq:FisherLocal}
\end{align}  
Using assumption A2 it easy to prove that $[J(\ve{\theta}^*)]_{ii}=\Ex_{\theta_i^*}\left[\frac{\partial^2\log p_i}{\partial\theta_i^2}\right]= - \Ex_{\theta_i^*}\left[\left(\frac{\partial\log p_i}{\partial\theta_i}\right)^2\right]=-[\tilde{\ve i}(\ve \theta^*)]_{ii}$. Finally, 
\begin{align*}
\sqrt{L}(\hat{\ve\theta}^L_{loc}-\ve\theta^*)\overset{a}{\sim} 
-J(\ve w^L)^{-1} \frac{1}{\sqrt{L}}\sum_{l=1}^{L} \ve\psi(\ve z_l;\ve\theta^*) \nonumber \\ \overset{a}{\sim} \N\left(\ve 0,\diag\left(\tilde{\ve i}(\ve \theta^*)\right)^{-1}\tilde{\ve i}(\ve \theta^*)\diag\left(\tilde{\ve i}(\ve \theta^*)\right)^{-1} \right)
\end{align*}  
Solving for $\hat{\ve\theta}^L_{loc}$ we obtain the following result
\begin{align}
\hat{\ve\theta}_{loc}^L \!\overset{a}{\sim}\!\left\{\begin{array}{ll}
\!\!\!\N\left(\ve 0,\frac{1}{L}\tilde{\ve i}^{-1}(\ve 0)\right) &\!\!\!\!\!\!\text{ under } \Hip_0\\
\!\!\!\N\left(\ve c,\frac{1}{L}\diag(\tilde{\ve i}(\ve c))^{-1}\tilde{\ve i}(\ve c)\diag(\tilde{\ve i}(\ve c))^{-1} \right) & \!\!\!\!\!\!\text{ under } \Hip_1
\end{array}\right.
\label{eq:lmle}
\end{align}

We now evaluate (\ref{eq:FisherLocal}) for the model (\ref{eq:htpeq}). The marginal distribution under $\Hip_1$ is $p_i(z_i(l);c_i) = \N(\sqrt{M} c_i,(1+c_i)^2)$. Given that the argument of the expectation only depends on $z_i(l)$ and $z_j(l)$, the expectation can be computed with respect to the jointly distribution on this pair of random variables 
\begin{align*}
& p_{ij}(z_i(l),z_j(l);c_i,c_j) = \\
& \N\left(
\sqrt{M}\left[\begin{array}{l}
c_i\\
c_j
\end{array}\right],
\left[\begin{array}{cc}
 (1+c_i)^2 & c_i c_j\\
 c_i c_j  &  (1+c_i)^2
\end{array}\right]
\right).
\end{align*}
Finally, making straightforward calculations we arrive to 
\begin{align*}
\left[\tilde{\ve i}(\ve c)\right]_{kj} 
& = \frac{\sigma_{kj}}{\sigma^2_k \sigma^2_j}\left(M+\frac{2\sigma_{kj}}{\sigma_k \sigma_j}\right) 
\end{align*}
where 
$\sigma_k=1+c_k$ and $\sigma_{kj}=c_k c_j + (1+2c_k)\delta_{kj}$.

\subsection{Asymptotic distribution of the local GLR}
Next we start with the proof of the asymptotic distribution of the local GLR. First, we recall that the \emph{global} MLE attains asymptotically the Cramer-Rao bound, i.e., it is asymptotically efficient, and therefore it satisfies:
\begin{equation}
\frac{\partial\log p(\ve z;\ve{\theta}^*)}{\partial \ve\theta} = L \ve i(\ve\theta^*)\left(\hat{\ve\theta}^L-\ve\theta^* \right)
\label{eq:CReq}
\end{equation}
where $\hat{\ve\theta}^L$ is the global MLE. We know that this estimator is consistent, i.e. $\hat{\ve\theta}^L\overset{p}{\rightarrow}\ve \theta^*$ as $L\rightarrow\infty$. As the local MLE is also consistent, we have that $\hat{\ve\theta}_{loc}^L\overset{p}{\rightarrow}\hat{\ve\theta}^L$ as $L\rightarrow\infty$. Thus, $(\ref{eq:CReq})$ is also satisfied with $\hat{\ve\theta}_{loc}^L$ instead of $\hat{\ve\theta}^L$ when $L\rightarrow\infty$. Then, using a first-order Taylor expansion of $\ve i(\ve{\theta}^*)$ around $\ve{\theta}^*$ and discarding the second order terms as $L\rightarrow\infty$, we have  
\begin{equation*}
\frac{\partial\log p(\ve z;\ve{\theta}^*)}{\partial \ve\theta} = L \ve i(\ve\theta^L_{loc})\left(\hat{\ve\theta}_{loc}^L-\ve\theta^* \right).
\end{equation*}
Integrating this equation with respect to $\ve \theta$ and evaluating at $\ve\theta=\ve\theta^*$:
\begin{align}
\log p(\ve z;\ve{\theta}^*) &= -\frac{L}{2} (\hat{\ve\theta}_{loc}^L-\ve\theta^* )^T \ve i(\ve\theta^L_{loc})(\hat{\ve\theta}_{loc}^L-\ve\theta^* ) \nonumber \\
&+ c(\hat{\ve\theta}_{loc}^L),
\label{eq:logp}
\end{align}
where the integration constant must be $c(\hat{\ve\theta}_{loc}^L) = \log p(\ve z;\hat{\ve\theta}_{loc}^L)$ given that (\ref{eq:logp}) is satisfied asymptotically by the consistence of $\hat{\ve\theta}_{loc}^L$ when $L\rightarrow\infty$. Therefore, 
\begin{equation*}
 p(\ve z;\ve{\theta}^*) = p\left(\ve z;\hat{\ve\theta}_{loc}^L\right) e^{-L\frac{1}{2} \left(\hat{\ve\theta}_{loc}^L-\ve\theta^* \right)^T \ve i(\ve\theta^L_{loc})\left(\hat{\ve\theta}_{loc}^L-\ve\theta^* \right)}.
\end{equation*}
Using the previous equation in the expression of local GLR, we obtain
\begin{align*}
T_{L}(\ve z)&=\frac{ p\left(\ve z;\hat{\ve\theta}_{loc}^L\right)}{ p\left(\ve z;\ve\theta^*=\ve 0\right)} = e^{L\frac{1}{2} \left(\hat{\ve\theta}_{loc}^L\right)^T \ve i(\ve\theta^L_{loc})\hat{\ve\theta}_{loc}^L },
\end{align*}
or 
\begin{align*}
2\log T_{L}(\ve\theta)&= L \left(\hat{\ve\theta}_{loc}^L \right)^T \ve i(\ve\theta^L_{loc})\hat{\ve\theta}_{loc}^L,
\end{align*}
which is the Wald test but using the local estimator instead of the global one.
Using again the CMT and the continuity of the second-order partial derivatives, the following is satisfied when $L\rightarrow\infty$:
\begin{equation*}
\ve i(\ve\theta^L_{loc})\hat{\ve\theta}_{loc}^L =\left\{
\begin{array}{l}
\ve i(\ve 0)\hat{\ve\theta}_{loc}^L \text{ under } \Hip_0\\
\ve i(\ve c)\hat{\ve\theta}_{loc}^L \text{ under } \Hip_1
\end{array}
\right.
\end{equation*}
Finally,
\begin{equation*}
2\log T_{L}(\ve\theta)\!=\!\left\{
\begin{array}{ll}
\!\!\!\! L (\hat{\ve\theta}_{loc}^L )^T \ve i(\ve 0)\hat{\ve\theta}_{loc}^L \overset{a}{\sim} \chi^2_{N} & \!\!\!\!\! \!\text{ under } \Hip_0\\
\!\!\!\! L (\hat{\ve\theta}_{loc}^L )^T \ve i(\ve c)\hat{\ve\theta}_{loc}^L \overset{a}{\sim} \chi^{'2}_{g,N} & \!\!\!\!\!\!\text{ under } \Hip_1
\end{array}
\right.
\end{equation*}
where, under $\Hip_0$ we use (\ref{eq:lmle}) and the fact that $\bar{\ve i}(\ve 0)=\ve i(\ve 0)=(M+2)I_N$. Under $\Hip_1$, we define the generalized chi-square distribution $\chi^{'2}_{g,N}$ with $N$ degrees of freedom as the distribution of the square norm of the Gaussian vector $\sqrt{L}\ve i(\ve c)^{\frac{1}{2}}\hat{\ve \theta}_{loc}$ with asymptotic distribution $\N\left(\sqrt{L}\ve i(\ve c)^{\frac{1}{2}}\ve c,\ve i(\ve c)^{\frac{1}{2}} \diag(\tilde{\ve i}(\ve c))^{-1} \tilde{\ve i}(\ve c) \diag(\tilde{\ve i}(\ve c))^{-1} \ve i(\ve c)^{\frac{1}{2}}\right)$.

So far, we have assumed nothing about the asymptotic behavior of the true parameter $\ve c$. In the particular case that there exists a constant $c_0$ such that $\|\ve c\|\! \leq \frac{c_0}{\sqrt{L}}$ (assumption A3), when $L\rightarrow\infty$ the covariance matrix of $\sqrt{L}\ve i(\ve c)^{\frac{1}{2}}\hat{\ve \theta}_{loc}$ becomes the identity matrix $\mathbf{I}_N$  and $\chi^{'2}_{g,N}$ becomes the non-central chi-square with $N$ degrees of freedom and non-centrality parameter $\lambda_{loc}\!=L(M+2)\|\ve c\|^2$, as is stated in (\ref{eq:lglr}).


\begin{thebibliography}{10}

\bibitem{shaikh2016energy}
Faisal~Karim Shaikh and Sherali Zeadally.
\newblock Energy harvesting in wireless sensor networks: A comprehensive
  review.
\newblock {\em Renewable and Sustainable Energy Reviews}, 55:1041--1054, 2016.

\bibitem{rashid2016applications}
Bushra Rashid and Mubashir~Husain Rehmani.
\newblock Applications of wireless sensor networks for urban areas: A survey.
\newblock {\em Journal of Network and Computer Applications}, 60:192--219,
  2016.

\bibitem{gungor2009industrial}
Vehbi~C Gungor, Gerhard~P Hancke, et~al.
\newblock Industrial wireless sensor networks: Challenges, design principles,
  and technical approaches.
\newblock {\em IEEE Trans. Industrial Electronics}, 56(10):4258--4265, 2009.

\bibitem{gubbi2013internet}
Jayavardhana Gubbi, Rajkumar Buyya, Slaven Marusic, and Marimuthu Palaniswami.
\newblock Internet of things (iot): A vision, architectural elements, and
  future directions.
\newblock {\em Future generation computer systems}, 29(7):1645--1660, 2013.

\bibitem{al2015internet}
Ala Al-Fuqaha, Mohsen Guizani, Mehdi Mohammadi, Mohammed Aledhari, and Moussa
  Ayyash.
\newblock Internet of things: A survey on enabling technologies, protocols, and
  applications.
\newblock {\em IEEE Communications Surveys \& Tutorials}, 17(4):2347--2376,
  2015.

\bibitem{chepuri2016sparse}
Sundeep~Prabhakar Chepuri and Geert Leus.
\newblock Sparse sensing for distributed detection.
\newblock {\em IEEE Transactions on Signal Processing}, 64(6):1446--1460, 2016.

\bibitem{ciuonzo2017distributed}
Domenico Ciuonzo and P~Salvo Rossi.
\newblock Distributed detection of a non-cooperative target via generalized
  locally-optimum approaches.
\newblock {\em Information Fusion}, 36:261--274, 2017.

\bibitem{al2018node}
Sara Al-Sayed, Jorge Plata-Chaves, Michael Muma, Marc Moonen, and Abdelhak~M
  Zoubir.
\newblock Node-specific diffusion lms-based distributed detection over adaptive
  networks.
\newblock {\em IEEE Transactions on Signal Processing}, 66(3):682--697, 2018.

\bibitem{tsitsiklis1989decentralized}
John~N Tsitsiklis et~al.
\newblock Decentralized detection.
\newblock 1989.

\bibitem{blum1997distributed}
Rick~S Blum, Saleem~A Kassam, and H~Vincent Poor.
\newblock Distributed detection with multiple sensors ii. advanced topics.
\newblock {\em Proceedings of the IEEE}, 85(1):64--79, 1997.

\bibitem{viswanathan1997distributed}
Ramanarayanan Viswanathan and Pramod~K Varshney.
\newblock Distributed detection with multiple sensors part i. fundamentals.
\newblock {\em Proceedings of the IEEE}, 85(1):54--63, 1997.

\bibitem{ChamberlandVeeravalli2003}
J.-F. Chamberland and V.V. Veeravalli.
\newblock {Decentralized detection in sensor networks}.
\newblock {\em {IEEE} Trans. Signal Process.}, 51(2):407--416, February 2003.

\bibitem{blum1996necessary}
Rick~S Blum.
\newblock Necessary conditions for optimum distributed sensor detectors under
  the neyman-pearson criterion.
\newblock {\em IEEE Transactions on Information Theory}, 42(3):990--994, 1996.

\bibitem{tenney1981detection}
Robert~R Tenney and Nils~R Sandell.
\newblock Detection with distributed sensors.
\newblock {\em IEEE Transactions on Aerospace and Electronic systems},
  (4):501--510, 1981.

\bibitem{willett2000good}
Peter Willett, Peter~F Swaszek, and Rick~S Blum.
\newblock The good, bad and ugly: distributed detection of a known signal in
  dependent gaussian noise.
\newblock {\em IEEE Transactions on Signal Processing}, 48(12):3266--3279,
  2000.

\bibitem{sayed2014adaptation}
Ali~H Sayed et~al.
\newblock Adaptation, learning, and optimization over networks.
\newblock {\em Foundations and Trends{\textregistered} in Machine Learning},
  7(4-5):311--801, 2014.

\bibitem{sayed2013diffusion}
Ali~H Sayed, Sheng-Yuan Tu, Jianshu Chen, Xiaochuan Zhao, and Zaid~J Towfic.
\newblock Diffusion strategies for adaptation and learning over networks: an
  examination of distributed strategies and network behavior.
\newblock {\em IEEE Signal Processing Magazine}, 30(3):155--171, 2013.

\bibitem{kar2008topology}
Soummya Kar, Saeed Aldosari, and Jos{\'e}~MF Moura.
\newblock Topology for distributed inference on graphs.
\newblock {\em IEEE Transactions on Signal Processing}, 56(6):2609--2613, 2008.

\bibitem{ciuonzo2014decision}
Domenico Ciuonzo and P~Salvo Rossi.
\newblock Decision fusion with unknown sensor detection probability.
\newblock {\em IEEE Signal Processing Letters}, 21(2):208--212, 2014.

\bibitem{hamed2012reliable}
S~Hamed Hamed and Ali Peiravi.
\newblock Reliable distributed detection in multi-hop clustered wireless sensor
  networks.
\newblock {\em IET Signal Processing}, 6(8):743--750, 2012.

\bibitem{varshney1986optimal}
Pramod~K Varshney et~al.
\newblock Optimal data fusion in multiple sensor detection systems.
\newblock {\em IEEE Transactions on Aerospace and Electronic Systems},
  (1):98--101, 1986.

\bibitem{kar2011distributed}
Soummya Kar, Ravi Tandon, H~Vincent Poor, and Shuguang Cui.
\newblock Distributed detection in noisy sensor networks.
\newblock In {\em Information Theory Proceedings (ISIT), 2011 IEEE
  International Symposium on}, pages 2856--2860. IEEE, 2011.

\bibitem{braca2010asymptotic}
Paolo Braca, Stefano Marano, Vincenzo Matta, and Peter Willett.
\newblock Asymptotic optimality of running consensus in testing binary
  hypotheses.
\newblock {\em IEEE Transactions on Signal Processing}, 58(2):814--825, 2010.

\bibitem{li2018fully}
Shang Li and Xiaodong Wang.
\newblock Fully distributed sequential hypothesis testing: Algorithms and
  asymptotic analyses.
\newblock {\em IEEE Transactions on Information Theory}, 64(4):2742--2758,
  2018.

\bibitem{cattivelli2011distributed}
Federico~S Cattivelli and Ali~H Sayed.
\newblock Distributed detection over adaptive networks using diffusion
  adaptation.
\newblock {\em IEEE Transactions on Signal Processing}, 59(5):1917--1932, 2011.

\bibitem{drakopoulos1991optimum}
Elias Drakopoulos and C-C Lee.
\newblock Optimum multisensor fusion of correlated local decisions.
\newblock {\em IEEE Transactions on Aerospace and Electronic Systems},
  27(4):593--606, 1991.

\bibitem{Tong_2007}
A.~Anandkumar, L.~Tong, and A.~Swami.
\newblock {Detection of Gauss-Markov Random Fields With Nearest-Neighbor
  Dependency}.
\newblock {\em {IEEE} Trans. Inf. Theory}, 55(2):816--827, Feb 2009.

\bibitem{anandkumar2009scalable}
Animashree Anandkumar.
\newblock Scalable algorithms for distributed statistical inference.
\newblock 2009.

\bibitem{javadi2016detection}
S~Hamed Javadi.
\newblock Detection over sensor networks: a tutorial.
\newblock {\em IEEE Aerospace and Electronic Systems Magazine}, 31(3):2--18,
  2016.

\bibitem{Lunden_Koivunen_Poor_2015}
J.~Lunden, V.~Koivunen, and H.~Poor.
\newblock Spectrum exploration and exploitation for cognitive radio: Recent
  advances.
\newblock {\em IEEE Signal Processing Magazine}, 32(3):123--140, May 2015.

\bibitem{Wang_Qiu_Zhao_2017}
B.~Wang, R.~C. Qiu, and Y.~Zhao.
\newblock Distributed source detection with dimension reduction in
  multiple-antenna wireless networks.
\newblock {\em IEEE Transactions on Vehicular Technology}, 66(4):2966--2980,
  Apr 2017.

\bibitem{Wang_Reiss_Cavallaro_2016}
L.~Wang, J.~D. Reiss, and A.~Cavallaro.
\newblock Over-determined source separation and localization using distributed
  microphones.
\newblock {\em IEEE/ACM Transactions on Audio, Speech, and Language
  Processing}, 24(9):1573--1588, Sep 2016.

\bibitem{Papoulis_Pillai_2002}
Athanasios Papoulis and S.~Unnikrishna Pillai.
\newblock {\em Probability, Random Variables and Stochastic Processes}.
\newblock McGraw-Hill Europe, 4th edition edition, Jan 2002.

\bibitem{Urkowitz_1967}
H.~Urkowitz.
\newblock Energy detection of unknown deterministic signals.
\newblock {\em Proceedings of the IEEE}, 55(4):523?531, Apr 1967.

\bibitem{Landau_Pollak_1962}
H.~J. Landau and H.~O. Pollak.
\newblock Prolate spheroidal wave functions, fourier analysis and uncertainty
  -- iii: The dimension of the space of essentially time- and band-limited
  signals.
\newblock {\em The Bell System Technical Journal}, 41(4):1295--1336, Jul 1962.

\bibitem{Slepian_1965}
David Slepian.
\newblock Some asymptotic expansions for prolate spheroidal wave functions.
\newblock {\em Journal of Mathematics and Physics}, 44(1--4):99--140, 1965.

\bibitem{Billingsley_1995}
Patrick Billingsley.
\newblock {\em Probability and Measure, 3rd Edition}.
\newblock Wiley-Interscience, 3 edition, Apr 1995.

\bibitem{Kay_SSP}
S.~Kay.
\newblock {\em Fundamentals of Statistical Signal Processing, Volume II:
  Detection Theory}.
\newblock Prentice-Hall, 1st ed. edition, 1998.

\bibitem{Levy_Det}
Bernard~C Levy.
\newblock {\em {Principles of signal detection and parameter estimation}}.
\newblock Springer, 2008.

\bibitem{Kay_SSP_ET}
Steven~M Kay.
\newblock Fundamentals of statistical signal processing, volume i: Estimation
  theory.
\newblock 1993.

\bibitem{White_1982}
Halbert White.
\newblock Maximum likelihood estimation of misspecified models.
\newblock {\em Econometrica}, 50(1):1--25, 1982.

\bibitem{Fortunati_Gini_Greco_Richmond_2017}
Stefano Fortunati, Fulvio Gini, Maria~S. Greco, and Christ~D. Richmond.
\newblock Performance bounds for parameter estimation under misspecified
  models: Fundamental findings and applications.
\newblock {\em IEEE Signal Processing Magazine}, 34(6):142--157, Nov 2017.

\bibitem{xiao2004fast}
Lin Xiao and Stephen Boyd.
\newblock Fast linear iterations for distributed averaging.
\newblock {\em Systems \& Control Letters}, 53(1):65--78, 2004.

\bibitem{cattivelli2010diffusion}
Federico~S Cattivelli and Ali~H Sayed.
\newblock Diffusion lms strategies for distributed estimation.
\newblock {\em IEEE Transactions on Signal Processing}, 58(3):1035--1048, 2010.

\bibitem{cabric2004implementation}
Danijela Cabric, Shridhar~Mubaraq Mishra, and Robert~W Brodersen.
\newblock Implementation issues in spectrum sensing for cognitive radios.
\newblock In {\em Signals, systems and computers, 2004. Conference record of
  the thirty-eighth Asilomar conference on}, volume~1, pages 772--776. Ieee,
  2004.

\bibitem{apostol1974mathematical}
Tom~M. Apostol.
\newblock {\em Mathematical analysis}, volume~2.
\newblock Addison-Wesley Reading, MA, 1974.

\bibitem{Cover_Thomas_2006}
Thomas~M. Cover and Joy~A. Thomas.
\newblock {\em Elements of Information Theory 2nd Edition}.
\newblock Wiley-Interscience, 2 edition, Jul 2006.

\bibitem{Vaart_2000}
A.~W. van~der Vaart.
\newblock {\em Asymptotic Statistics}.
\newblock Cambridge University Press, Jun 2000.

\end{thebibliography}

\end{document}